\documentclass{aa}  
\usepackage[varg]{txfonts}
\usepackage{graphicx}
\usepackage{siunitx}
\usepackage{xcolor}
\usepackage{natbib}
\usepackage{booktabs}
\usepackage{nccmath}
\usepackage{amsmath}
\usepackage{tabularx}
\usepackage{algorithm2e}
\usepackage{csquotes}
\usepackage{comment}
\usepackage{gensymb}
\usepackage{enumitem}
\usepackage{subcaption}
\usepackage[labelfont=bf]{caption}
\usepackage{mwe}
\usepackage[colorlinks,citecolor=blue,urlcolor=blue,bookmarks=false,allcolors=blue,hypertexnames=true]{hyperref} 
\usepackage{threeparttable}
\usepackage{color}
\usepackage[colorinlistoftodos]{todonotes}

\def \kms         {km$\,$s$^\mathrm{-1}$}

\def \deg         {\text{$^\mathrm{\circ}$}}
\def \arcmin      {\text{$^\prime$}}
\def \arcsec      {\text{$^\mathrm{\prime\prime}$}}

\def \mjybeam     {mJy\,beam$^\mathrm{-1}$}

\newlist{inlineroman}{enumerate*}{1}
\setlist[inlineroman]{itemjoin*={{ }},afterlabel=~,label=\roman*)}

\makeatletter
\renewcommand{\fnum@figure}{Figure \thefigure}
\makeatother

\begin{document} 
\title{Unmasking the history of 3C\,293 with LOFAR sub-arcsecond imaging \thanks{Fits files of the radio maps are available at the CDS via anonymous ftp to cdsarc.u-strasbg.fr (130.79.128.5)
or via http://cdsweb.u-strasbg.fr/cgi-bin/qcat?J/A+A/}}
\author{Pranav Kukreti\inst{1,2}\thanks{\email{kukreti@astro.rug.nl}},
Raffaella Morganti\inst{2,1}, Timothy W. Shimwell\inst{2,3}, Leah K. Morabito\inst{4,5}, Robert J. Beswick\inst{6}, Marisa Brienza\inst{7,8}, Martin J. Hardcastle\inst{9}, Frits Sweijen\inst{3}, Neal Jackson\inst{6}, George K. Miley\inst{3}, Javier Moldon\inst{10}, Tom Oosterloo\inst{2,1}, Francesco de Gasperin\inst{11,8}}

\authorrunning{Kukreti et al.}
\titlerunning{Unmasking the history of 3C\,293}

\institute{
Kapteyn Astronomical Institute, University of Groningen, Postbus 800, 9700 AV Groningen, The Netherlands
\and
ASTRON, the Netherlands Institute for Radio Astronomy, Oude Hoogeveensedijk 4, 7991 PD Dwingeloo, The Netherlands
\and
Leiden Observatory, Leiden University, P.O.Box 9513, NL-2300 RA, Leiden, The Netherlands
\and
Centre for Extragalactic Astronomy, Department of Physics, Durham University, Durham DH1 3LE, UK
\and
Institute for Computational Cosmology, Department of Physics, University of Durham, South Road, Durham DH1 3LE, UK
\and
Jodrell Bank Centre for Astrophysics, School of Physics and Astronomy, University of Manchester, Oxford Rd, Manchester M13 9PL,UK
\and
Dipartimento di Fisica e Astronomia, Università di Bologna, via P. Gobetti 93/2, 40129, Bologna, Italy
\and
INAF - Istituto di Radioastronomia, Via P. Gobetti 101, 40129, Bologna, Italy
\and
Centre for Astrophysics Research, University of Hertfordshire, College Lane, Hatfield AL10 9AB, UK
\and
Instituto de Astrof\'isica de Andaluc\'ia (IAA, CSIC), Glorieta de las Astronom\'ia, s/n, E-18008 Granada, Spain
\and
Hamburger Sternwarte, Universit\"at Hamburg, Gojenbergsweg 112, D-21029, Hamburg, Germany
}

  \abstract
   {Active galactic nuclei (AGNs) show episodic activity, which can be evident in galaxies that exhibit restarted radio jets. These restarted jets can interact with their environment, leaving signatures on the radio spectral energy distribution. Tracing these signatures is a powerful way to explore the life of radio galaxies. This requires resolved spectral index measurements over a broad frequency range including low frequencies. We present such a study for the radio galaxy 3C\,293, which has long been thought to be a restarted galaxy on the basis of its radio morphology.
   Using the International LOFAR telescope (ILT) we probed spatial scales as fine as $\sim$0.2\arcsec at 144\,MHz, and to constrain the spectrum we combined these data with Multi-Element Radio Linked Interferometer Network (MERLIN) and Very Large Array (VLA) archival data at frequencies up to 8.4\,GHz that have a comparable resolution. 
   In the inner lobes ($\sim$2\,kpc), we detect the presence of a spectral turnover that peaks at $\sim$225\,MHz and is most likely caused by free-free absorption from the rich surrounding medium. We confirm that these inner lobes are part of a jet-dominated young radio source (spectral age $\lesssim$0.17 Myr), which is strongly interacting with the rich interstellar medium (ISM) of the host galaxy. The diffuse emission surrounding these lobes on scales of up to $\sim$4.5\,kpc shows steeper spectral indices ($\Delta\alpha\sim$ 0.2-0.5, S $\propto \nu^{-\alpha}$) and a spectral age of $\lesssim$0.27\,
   Myr. The outer lobes (extending up to $\sim$100\,kpc) have a spectral index of $\alpha$ $\sim$ 0.6-0.8 from 144-4850\,MHz with a remarkably uniform spatial distribution and only mild spectral curvature ($\Delta\alpha\lesssim$ 0.2). We propose that intermittent fuelling and jet flow disruptions are powering the mechanisms that keep the spectral index in the outer lobes from steepening and maintain the spatial uniformity of the spectral index. Overall, it appears that 3C\,293 has gone through multiple (two to three) epochs of activity. This study adds 3C\,293 to the new sub-group of restarted galaxies with short interruption time periods. This is the first time a spatially resolved study has been performed that simultaneously studies a young source as well as the older outer lobes at such low frequencies. This illustrates the potential of the International LOFAR telescope to expand such studies to a larger sample of radio galaxies.}

    \keywords{galaxies: active – radio continuum: galaxies – galaxies: individual: 3C\,293 - techniques: high angular resolution}

    \date{Received March 15, 2021; accepted July 01, 2021.}    

    \maketitle
\section{Introduction}

Over the past decades, active galactic nuclei (AGNs) have been demonstrated to show episodic activity and have been identified in different phases of their life cycle. Lobes of remnant plasma from a previous phase of activity coexisting with a newly born pair of radio jets are typical indicators of restarted or episodic activity in such galaxies (for a review, see \citealt{Saikia2009}). Restarted radio galaxies have been used to constrain the timescales of activity and quiescence (and therefore the duty cycle), which are crucial to understand the life cycle of these galaxies \citep{Morganti2017b}.\par
The life cycle of a radio galaxy is understood to start from a phase of morphologically compact radio emission with an absorbed or steep spectrum. Such sources are called compact steep spectrum (CSS) and gigahertz peaked spectrum (GPS) sources \citep{ODea1998b, Fanti2009,ODea2021a}. These sources show morphological similarities to large-scale radio sources, but with sizes of just a few kiloparsecs (similar to galactic scales), and they are thought to develop into large-scale radio galaxies. However, intermittent AGN activity may prevent some of these types of sources from growing to large-scale radio galaxies.
\par
After an initial phase of activity that can last between $\sim$10$^\mathrm{7}$ yr and $\sim$10$^\mathrm{8}$ yr \citep{Parma1999,Parma2007,Hardcastle2018}, the nuclear activity stops and the injection of fresh plasma to the lobes ceases. 
In some cases, the cessation of activity can lead to observable remnant plasma lobes without any activity near the core \citep{Parma2007,Murgia2010a}. These lobes are heavily affected by radiative losses that causes spectral steepening. This remnant phase has been detected in a small fraction (<10\%) of radio galaxies \citep{Saripalli2012b,Brienza2017,Mahatma2018,Quici2021b}. A slightly more frequently observed (13-15\% of radio galaxies; \citealt{Jurlin2020}) scenario is one where the activity is intermittent and radio plasma from an older phase as well as radio jets from a newer phase are simultaneously visible suggesting that activity has restarted after a relatively short remnant phase \citep{Schoenmakers1998a,Stanghellini2005,Shulevski2012a}. 

Indeed, the results from the growing statistics of radio galaxies in less common phases, such as remnants and restarted galaxies, combined with new modelling confirm not only the presence of a life cycle of activity, but also favour a power-law distribution for the ages, which implies a high fraction of short-lived AGNs (see \citealt{Shabala2020a,Morganti2020}).
\par 
Restarted radio galaxies provide a unique opportunity to observe and subsequently model the spectral and morphological properties of the older and newer lobes simultaneously. The characterisation of these properties can be used to estimate ages and timescales of their period of activity. \par 
Identifying restarted galaxies is not easy as they can show a variety of properties depending on the radio galaxy itself. A well known group of restarted galaxies are the double-double radio galaxies (DDRGs; \citealt{Schoenmakers2000}) in which a new pair of radio lobes are seen closer to the nucleus than the older lobes. Restarted galaxies with three pairs of radio lobes have also been identified, the so-called 'triple-double' galaxies, for example B\,0925+420 \citep{Brocksopp2007}, 'Speca' \citep{Hota2011} and J\,1216+0709 \citep{Singh2016}. In the case of DDRGs, spectral age studies have provided estimates of the timescale of the quiescent phase to be between 10$^\mathrm{5}$ yr and 10$^\mathrm{7}$ yr and at most, $\sim$50\% of the length of the previous active phase \citep{Konar2013EpisodicActivity,Orru2015Wide-fieldDoubeltjes,Nandi2019,Marecki2020,Marecki2021a}.\par
DDRGs represent only a fraction of restarted galaxies and in other cases, compact inner jets are found embedded in low-surface brightness, large-scale lobes \citep{Jamrozy2009, Kuzmicz2017OpticalActivity} or a relatively bright core \citep{Jurlin2020}. However, it is not always possible to identify restarted galaxies based on morphology alone. Over the years, spectral properties of radio galaxies have also been used to identify restarted radio galaxies \citep{Parma2007,Murgia2010a,Jurlin2020,Morganti2020}. One such interesting case is of 3C\,388, where a dichotomy in the spectral index distribution between different regions of the lobes was found which indicated two different jet episodes \citep{Burns1982,Roettiger1994}. More recently, \citealt{Brienza2020} used the LOw Frequency Array (LOFAR) \citep{VanHaarlem2013}
to confirm the presence of restarted activity in this galaxy. \citet{Jurlin2020} used the presence of a steep spectrum core ($\alpha^\mathrm{150\,MHz}_\mathrm{1360\,MHz} > 0.7$, where S$_\mathrm{\nu}\propto \nu^\mathrm{-\alpha}$), along with morphological properties such as low surface brightness extended emission to identify candidate restarted galaxies. \par

Over the past few years, a new sub-group of candidate restarted radio galaxies has been found, which do not show spatially resolved spectral properties expected from old remnant plasma, for example an ultra steep spectrum ($\alpha>1.2$) and a steep curvature in the radio spectra ($\Delta\alpha\geq 0.5$). Instead, these sources show bright inner jets and very diffuse outer lobes with a homogeneous spatial distribution of spectral index, for example Centaurus A \citep{McKinley2018,McKinley2013,Morganti1999},  B2~0258+35 \citep{Brienza2018}, and more recently NGC\,3998 \citep{Sridhar2020}. Although the physical mechanism responsible for these properties is still under debate, some studies indicate that the older outer lobes could still be fuelled at low levels by the active inner jets \citep{McKinley2018}. This is different from the model of typical restarted galaxies where fuelling of the older lobes has stopped. Therefore, this sub-group poses a new challenge to our understanding of the life-cycle of radio galaxies.\par 
Characterising the spectral properties of restarted galaxies is challenging, because it requires a wide frequency coverage to frequencies above 1.4\,GHz to include the high frequency end, where the effects of radiative losses are dominant due to ageing and cessation of fuelling, and down to a few tens/hundreds of\,MHz, where the signatures of plasma injection are alive for the longest time. One way to study such galaxies is to use the International LOFAR Telescope (ILT) which allows us to resolve the emission from such galaxies at arcsecond resolution down to low frequencies (144\,MHz), and complement observations at GHz frequencies. The LOFAR telescope includes 13 international stations, providing baselines of up to 1989 km which translates to an angular resolution of 0.27\arcsec at 150\,MHz. Although these international stations have always been available, there have only been a handful of sub-arcsecond studies with the telescope \citep{Moldon2014,Jackson2016,Morabito2016,Varenius2015,Varenius2016,Ramirez-Olivencia2018,Harris2019a,Kappes2019}. This is mainly due to the fact that the calibration with the full international LOFAR telescope is technically challenging. However, with the development of a new calibration strategy, presented in detail in \citet{Morabito2021}, it is now possible to more routinely perform high resolution low frequency studies of these galaxies. We can make use of this capability to search for the presence of a newer phase of activity in restarted radio galaxies and study the properties of small-scale (a few kpc) emission down to\,MHz frequencies. Combined with the Dutch stations of LOFAR, we can also study the large-scale emission (hundreds of kpc) from these galaxies. 
\par

This paper is structured in the following manner: In Sect. 1, we give an introduction to the source, in Sect. 2, we describe the data and the data reduction procedures, and the procedure to make spectral index maps; in Sect. 3 we present the large and small-scale source morphology, the large and small-scale spectral index, large-scale spectral age analysis, and the absorption models for the inner lobes in the centre. In Sect. 4, we discuss the properties and results for the central region and the outer lobes and then summarise the evolutionary history of 3C\,293. 
Throughout the paper, we define the spectral index $\alpha$, was calculated using the convention: $S\propto\nu^\mathrm{-\alpha}$. The cosmology adopted in this work assumes a flat universe with H$_\mathrm{0}$=71 km s$^\mathrm{-1}$ Mpc$^\mathrm{-1}$, $\Omega_\mathrm{m}=0.27$ and $\Omega_\mathrm{vac}=0.73$. At the redshift of 3C\,293, 1\arcsec corresponds to 0.873 kpc.
\section{Overview of the source 3C\,293}
\begin{figure*}
        \centering
        \begin{subfigure}[b]{0.49\textwidth}  
            \centering 
            \includegraphics[width=0.9\textwidth]{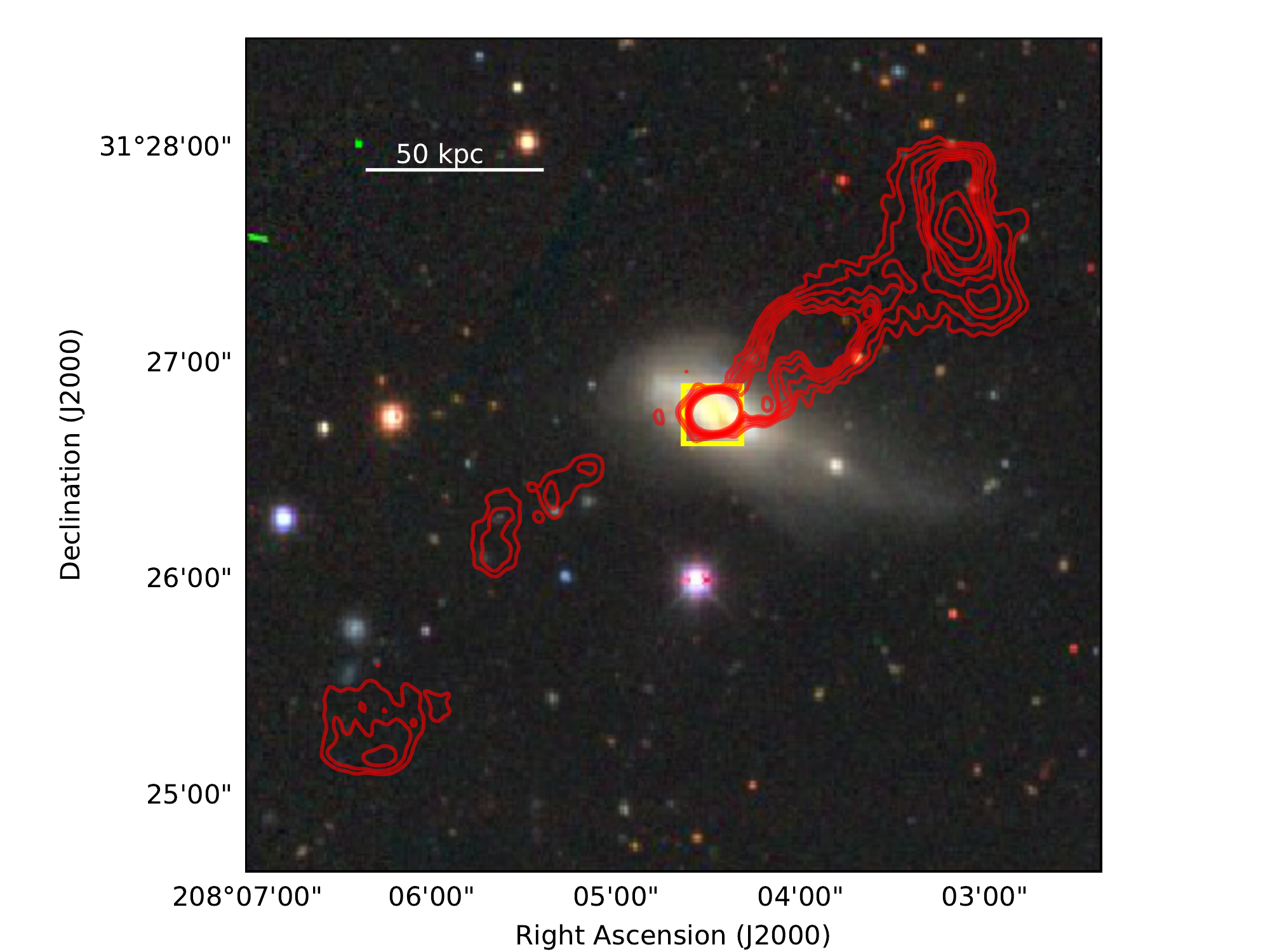}
            \caption[2]%
            {}  
            \label{fig:opt_largescale}
        \end{subfigure}     
        \begin{subfigure}[b]{0.49\textwidth}
            \centering
            \includegraphics[width=\textwidth]{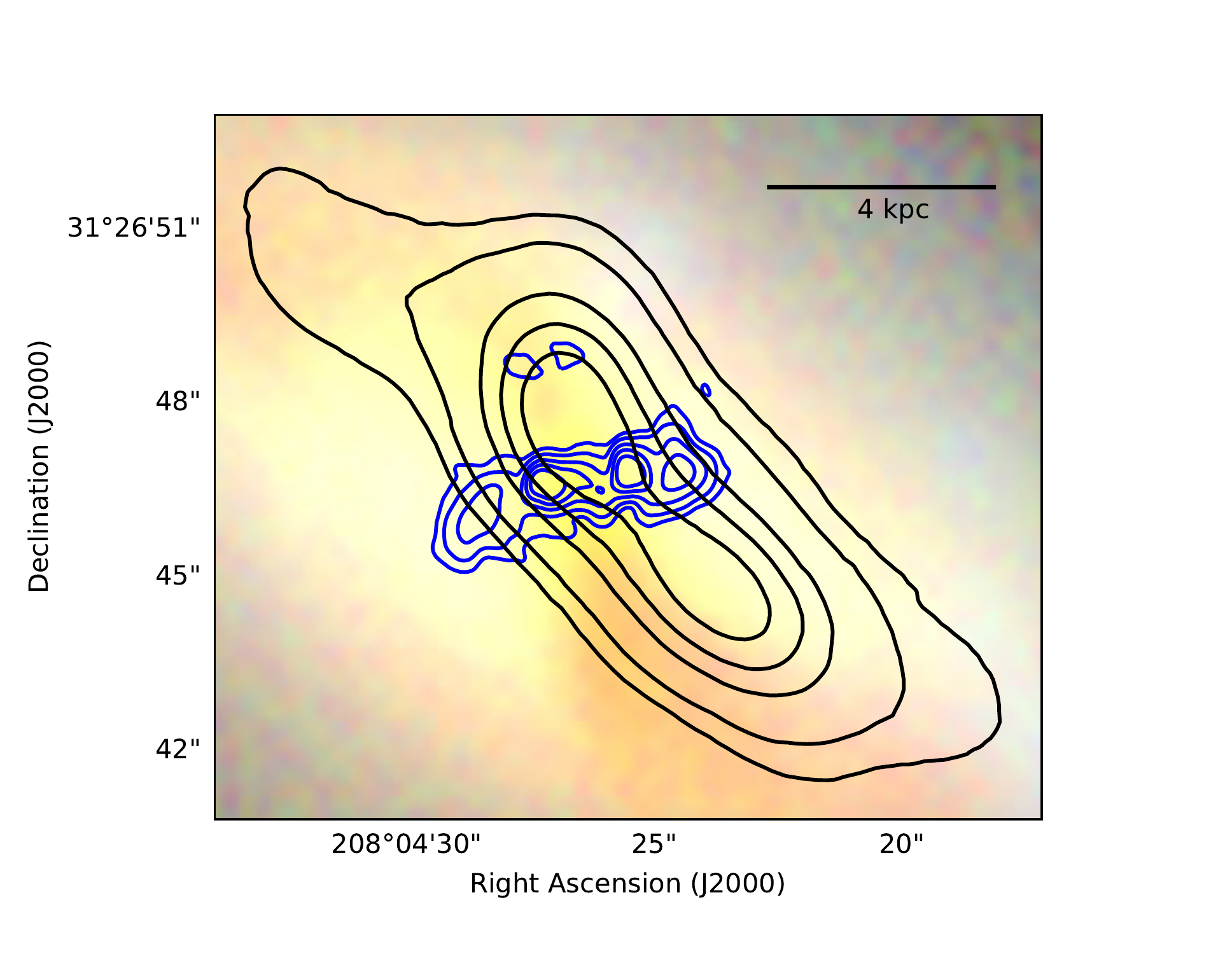}
            \caption[1]%
            {}    
            \label{fig:opt_smallscale}
        \end{subfigure}
        \caption[optical]   
        {\small (a) RGB colour image of the host galaxy of 3C\,293 from DECaLS\footnote{\url{https://www.legacysurvey.org/}} \citep{Dey2019}. A small companion galaxy can be seen in the south-west, $\sim$37\arcsec away from the host galaxy. The red contours show the Very Large Array (VLA) 1.36\,GHz emission at 4.6\arcsec$\times$4.1\arcsec resolution from archival data we have reprocessed. The outer lobes extend in the north-west and south-east direction and are aligned almost perpendicular to the orientation of the dust lanes and the gas disk. The blue box marks the region covered by the blue contours in the right panel. (b) DECaLS RGB colour image overlaid with black contours that map the CO(1-0) emission disc from \citet{Labiano2014} with the south-western side approaching us and northeastern side receding. The blue contours show the 120-168\,MHz radio continuum emission from the inner $\sim$4.5\,kpc as seen in our new LOFAR image. The inner lobes extend in east and west and can be seen to bend out of the host galaxy's CO disk, especially in the east.} 
        \label{optical}
    \end{figure*}

3C\,293 is a nearby radio galaxy at a redshift of $z = 0.045$ \citep{DeVaucouleurs1991}. The large and small-scale structure of 3C\,293 has been studied before in the radio and has a number of peculiarities. When observed at arcsec-resolution the source shows two asymmetric radio lobes and a central compact component (Figure~\ref{fig:opt_largescale}) and a total extension of $\sim$220\,kpc.  \citet{Bridle1981} were the first to study the asymmetric large-scale outer lobes that have bright concentrations of emission at their further points from the core. A total luminosity of L(1.4\,GHz) = 2$\times10^\mathrm{25}$ W Hz$^\mathrm{-1}$ places 3C\,293 on the border of the FR-I/FR-II classification \citep{Fanaroff1974}. The bright steep-spectrum centre and presence of large and small-scale lobes have been used to identify 3C\,293 as a candidate for restarted activity \citep{Bridle1981,Akujor1996b}.\par
Sub-arcsecond resolution studies have resolved the $\sim$4.5\,kpc central region, shown in Figure~\ref{fig:opt_smallscale}. Two prominent peaks of emission, understood to be the inner lobes, were found embedded in diffuse emission \citep{Akujor1996b,Beswick2002,Beswick2004}. There is an abrupt drop in the surface brightness of the outer lobes, which is $\sim$2500 times lower compared to the diffuse emission in the centre. \citet{Beswick2004} and \citet{Floyd2005} concluded that the eastern lobe is approaching us and the western lobe is receding. This picture of the orientation was also suggested by \citet{Mahony2016} (see their Figure 7) who studied the ionised gas outflows in the inner few kpc. One of the most striking features of the radio morphology of the galaxy is the $\sim$35$\deg$ (projected) misalignment between the inner and outer lobes, the origin of which is still unclear, although \citet{Bridle1981} suggested that the misalignment could be explained by radio jet refraction due to pressure gradients in dense circumgalactic atmospheres. More recently, \citet{Machalski2016a} concluded that a fast realignment of the jet axis, resulting from a rapid flip of the black hole spin, could be responsible for this misalignment. Such misalignment is rare in radio galaxies.\par
\citet{Joshi2011} performed an integrated spectral index study of the source covering a frequency range of 154\,MHz to 4860\,MHz by using the Giant Metrewave Radio Telescope (GMRT) and the VLA and estimated spectral ages. They obtained a straight spectrum for several regions of the source with a spectral index of $\alpha^{154}_{4860}$ = 0.72$\pm$0.02, 0.80$\pm$0.02 and 0.91$\pm$0.03 for the central region, outer north-western lobe and outer south-eastern lobe, respectively. Assuming an equipartition magnetic field and a break frequency equal to the highest frequency of their observations, they derived a spectral age of $\leq$16.9 Myr and $\leq$23 Myr for the north-western and south-eastern lobes respectively. Using a jet speed of $\mathrm{c}$ and a hotspot lifetime of $\sim 10^\mathrm{4}-10^\mathrm{5}$yr, they estimated a time period of $\sim 0.1$Myr for the interruption of activity. Since the real jet speed will be lower than $\mathrm{c}$ ($\lesssim$0.3-0.6$\mathrm{c}$; \citealt{Arshakian2004a,Jetha2006}), the actual interruption time period is of course, higher. They argue that within this time the inner double must also form. However, the integrated spectral index measurements for the outer lobes can not give any information about the spectral properties of plasma as a function of distance from the centre, needed to understand the physical processes active in the lobe. Detailed resolved studies are required for this purpose.\par
From studies at other wavelengths we know that 3C\,293 is hosted by a peculiar elliptical galaxy VV5-33-12 showing multiple dust lanes and compact knots \citep{VanBreugel1984,Martel1999,DeKoff2000,Capetti2000}. An RGB colour image of the host galaxy is shown in Figure~\ref{optical}. The host galaxy is a post-coalescent merger that is undergoing a minor interaction with a close satellite galaxy that lies to $\sim$37\arcsec towards the south-west \citep{Emonts2016a}. \citealt{Emonts2016a} suggest that the merger in 3C\,293 may provide the fuel to trigger the AGN activity. Large amounts of gas (M(cold H$_\mathrm{2}$) = 2.3$\times10^\mathrm{10}$ M$_\mathrm{\odot}$) = 3.7$\times10^\mathrm{9}$ M$_\mathrm{\odot}$) have also been found in both emission and absorption in the central few kiloparsecs \citep{Evans1998,Evans2005,Ogle2010} confirming the presence of a dense ISM. \citet{Labiano2014} find that the molecular gas (traced by CO(1-0)) is distributed along a 21\,kpc diameter warped disk, that rotates around the AGN (Figure~\ref{fig:opt_smallscale}).  \par

\par
\section{Observations and data reduction}

\begin{figure*}
        \centering
        \begin{subfigure}{0.7\textwidth}
            \centering
            \includegraphics[width=\textwidth]{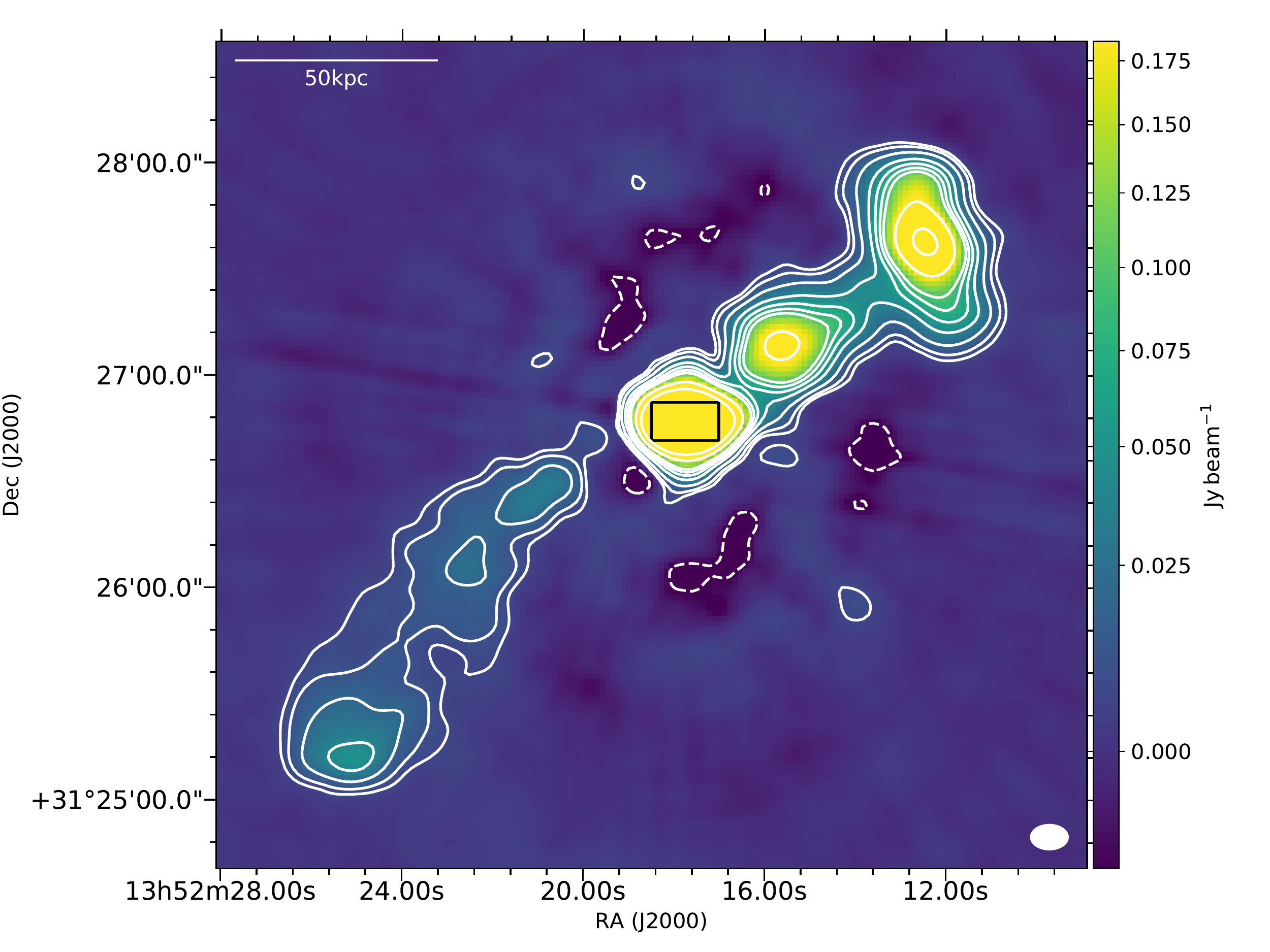}
            \caption[1]%
            {Dutch array}    
            \label{fig:hba_low}
        \end{subfigure}
        \begin{subfigure}{0.7\textwidth}  
            \centering 
            \includegraphics[width=\textwidth]{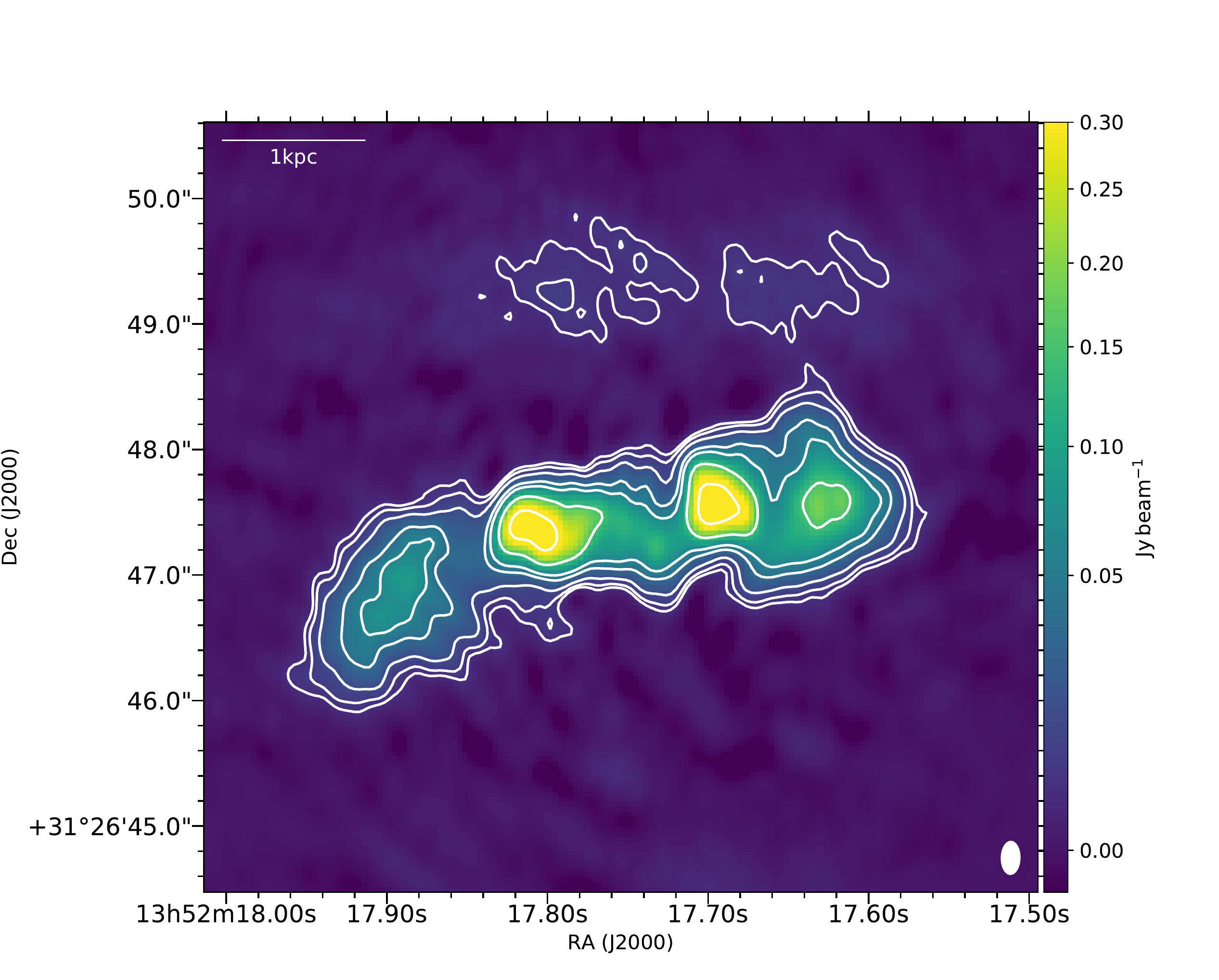}
            \caption[2]%
            {International array}  
            \label{fig:hba_high}
        \end{subfigure}     
        
        \caption[international]   
        {\small (a) LOFAR HBA image with Dutch array from LoTSS-DR2 data showing the large-scale ($\sim$250\arcsec) structure in 3C\,293 at 144\,MHz with a resolution of 10.5\arcsec$\times$7\arcsec. The contours marking the large-scale structure are at: (-3, 3, 5, 10, 20, 30, 40, 50, 100, 200)$\times\sigma_\mathrm{RMS}$ where $\sigma_\mathrm{RMS}$ = 2 \mjybeam is the RMS noise in the image. The black rectangle in the centre marks the region that is shown in the high resolution image. (b) The LOFAR HBA image with international stations at 144\,MHz of the central 4.5\,kpc region with a resolution of 0.26\arcsec$\times$0.15\arcsec. The small-scale inner lobes and diffuse emission are visible and are marked with contours at: (-3, 3, 5, 10, 25, 40, 100, 250)$\times\sigma_\mathrm{RMS}$ where $\sigma_\mathrm{RMS}$ = 1.5 \mjybeam is the RMS noise in the image. Image statistics for all images are summarised in Table~\ref{image_data}       } 
        \label{lofar_hba}
    \end{figure*}
\begin{table*}
\centering
\begin{threeparttable}
\caption[]{Summary of LOFAR observations.}
         \label{observation_data}
\begin{tabular}{cccccccc}
            \hline
            \hline
    \noalign{\smallskip}
    Configuration  & Central frequency & Bandwidth & TOS$^\mathrm{1}$ & Interval$^\mathrm{2}$ & Calibrators & Observation date\\
     & (MHz) & (MHz) & (hours) & (seconds) & & &\\
    \noalign{\smallskip}
    \hline
    \hline
    \noalign{\smallskip}
    HBA Dutch$^\mathrm{3}$ & 144  & 48 & 8 & 1 & 3C196, 3C295 & 12 April 2018 \\
    HBA International & 144  & 48 & 4+4 & 1 & 3C295 & 30 July and 02 August 2020 \\
            \hline
            \hline
 \end{tabular}      
      \small{$^\mathrm{1}$ Time on source for the target, $^\mathrm{2}$ Sample integration interval (time resolution of the data), $^\mathrm{3}$ HBA Dutch array data from LoTSS-DR2}
\end{threeparttable} 
\end{table*}
In order to trace the radio emission and characterise the spectral properties over a broad range of frequencies and on large and small scales, we used data from different telescopes. We use new observations with the ILT High Band Antenna (HBA) at 120-168\,MHz. We combine these with archival VLA data at 1.36\,GHz and 4.85\,GHz for a low resolution spectral study. For the high resolution complementary data, we use archival Multi-Element Radio Linked Interferometer Network (MERLIN) 1.36\,GHz, VLA 4.85\,GHz and 8.45\,GHz data.

\subsection{LOFAR HBA observations}

We performed targeted observations of 3C\,293 with the ILT HBA for a total of 8 hours split into two 4 hour observing runs, on 30 July 2020 and 02 August 2020 (Project code - LC14$\_$015). The 60 Dutch stations and 12 international stations were used as a part of the ILT array.
The international stations used for the observations were - DE601, DE602, DE603, DE604, DE605, FR606, SE607, UK608, DE609, PL610, PL612 and IE613.
The observations were carried out with the standard survey setup \citep{Shimwell2019b}, with a 48\,MHz bandwidth centred at 144\,MHz. The bandwidth was split into channels of 12.2 KHz width with an integration time of 1s. 3C\,295 and 3C\,196 were used as the flux density calibrators and observed for 10 mins before and after each target observation. After observing, the data was passed through the standard LOFAR pre-processing pipeline \citep{Heald2010a} where the RFI flagging (AOFlagger; \citealt{Offringa2010Post-correlationMethods,Offringa2012ADetection}) and averaging down to 1 second per sample and 4 channels per subband, was carried out. The processed measurement sets were then passed through the PREFACTOR\footnote{\url{https://github.com/lofar-astron/prefactor}} pipeline \citep{DeGasperin2019b,VanWeeren2016,Williams2016LOFARCounts} with the default parameter settings for direction independent calibration. A high resolution model of 3C\,295 was used for calibration.
Stations RS509 and PL610 were flagged for the 02 August data set due to bad data.\par
The data from the Dutch and international array have to be reduced using different procedures. This is due to fact that the international stations have different clocks and beams than the core and remote stations of ILT. Also, at low frequencies, the ionospheric effects become immensely relevant where they play an important role in corrupting astronomical signals \citep{Intema2009}. The wide geographical spread of the ILT means that different international stations see through very different regions of atmosphere. The data reduction procedures are described in the next two subsections.

\begin{table*}
\centering
\begin{threeparttable}
\caption[]{Image statistics summary.}
         \label{image_data}
\begin{tabular}{cccccc}
            \hline
            \hline
    \noalign{\smallskip}
    Frequency & Telescope  & Resolution (BPA) & RMS noise  & Integrated flux density $\pm$ error\\
    (MHz) & & (\arcsec) & (\mjybeam) & (Jy)\\
    \noalign{\smallskip}
    \hline
    \hline
    \noalign{\smallskip}
    144 & LOFAR HBA Dutch & 10.5\arcsec$\times$7\arcsec  (90\degree) & 2.0 & 15.19$\pm$1.52 \\
    1360 & VLA B configuration & 4.6\arcsec$\times$4.1\arcsec  (74\degree) &0.08 & 4.68$\pm$0.23 \\
         \vspace*{10px}
    4850 & VLA C configuration & 6.0\arcsec$\times$6.0\arcsec (90\degree) & 0.22 & 1.88$\pm$0.09 \\
    144 & LOFAR HBA international & 0.26\arcsec$\times$0.15\arcsec (1\degree) & 1.5 & 10.04$\pm$1.01 \\
    1360 & MERLIN & 0.24\arcsec$\times$0.21\arcsec (33\degree) & 1.2 & 3.72$\pm$0.19 \\ 
    4850 & VLA C configuration & 0.36\arcsec$\times$0.30\arcsec (53\degree) & 0.2 & 1.49$\pm$0.07 \\
    8450 & VLA C configuration & 0.30\arcsec$\times$0.26\arcsec (70\degree) & 0.26 & 1.01$\pm$0.05 \\
            \hline
            \hline
 \end{tabular}      
    \small{The first three rows show measurements of the entire $\sim$200\,kpc source (centre+outer lobes) from the low resolution images. The last four rows show measurements from the high resolution image of the $\sim$4.5\,kpc centre.}
\end{threeparttable} 
\end{table*}
\subsubsection{High resolution LOFAR International array}
The processed data from the PREFACTOR pipeline, including the international stations, were averaged and calibrated using the LOFAR long baseline pipeline\footnote{\url{https://github.com/lmorabit/lofar-vlbi}} \citep{Morabito2021}, which performs in-field delay calibration using a bright calibrator in the field of view and cross-matching it with the LOFAR Long-Baseline Calibrator Survey \citep{Jackson2016,Jackson2021} and LOFAR Two-metre Sky Survey (LoTSS) \citep{Shimwell2017}. The pipeline first averages the data to a resolution of 8s in time and 97.64 KHz in frequency, that is two channels per subband. It then performs dispersive delay calibration using the in-field delay calibrator close to the target, which in our case was L465974 (J2000.0 RA 13$^{h}$53$^{m}$11.69$^{s}$, Dec. +32\degree05\arcmin42.6\arcsec), located at an angular distance of 0.7$\degree$ from the target. The two 4\,hr data sets were processed separately and were combined after applying their respective delay calibration solutions. The combined 8\,hr data set was then used for self-calibration and imaging of the delay calibrator using the procedure outlined in \citet{VanWeeren2020} which makes use of WSClean \citep{Offringa2014Wsclean:Astronomy,Offringa2017AnImages} for imaging. We estimated an integrated flux density of 0.78\,Jy for the delay calibrator, which is within 10\% of the integrated flux density of 0.85\,Jy from LoTSS-DR2. This gives us confidence in our flux scale.\par 

The dispersive delay calibrator solutions were then transferred to the target separately for each 4\,hr data set. Rounds of self-calibration and imaging were performed on the combined calibrated target data set using the same procedure with a Briggs weighting scheme and a robust of $-$1 which gave a synthesised beamwidth of 0.26\arcsec$\times$0.15\arcsec. 
We measure an integrated flux density of 10.04$\pm1.01$\,Jy for the target, which is in good agreement with the integrated flux density of the central region which is 10.70$\pm$1.07\,Jy (see region $C$ in Table~\ref{region_summary}). The RMS noise in the final map is $\sim$0.2 \mjybeam which increases up to $\sim$1.5 \mjybeam  near the target. The thermal noise for the image is 0.08-0.1 \mjybeam and the noise level in our image is dominated presumably by residual phase and amplitude errors around a bright source. We have used a flux scale error of 10\% and the final high resolution map is shown in Figure~\ref{fig:hba_high}.
 
\subsubsection{LOFAR Dutch array and LoTSS-DR2 resolution}
The quality of the low resolution image made with only the Dutch array of the targeted observations was not good enough to perform a spectral analysis, compared to a similar resolution image made with a LoTSS-DR2 (Shimwell et al. in prep) pointing. Hence its data reduction process is not discussed further.\par 
We have instead used the Dutch array data from a LoTSS-DR2 pointing with the reference code P207+32 (Project code - LC7$\_$024) for the low resolution image. In this pointing, the target lies 1.2$\degree$ away from the phase centre. The observations were carried out using the standard survey setup, that is 8 hours on-source time, 48\,MHz bandwidth centred at 144\,MHz divided over 231 subbands, and 1 second integration time. The Dutch array data are averaged to 8s in time and 2 channels per 1.95\,MHz subband. The flux density calibrator was 3C\,196. Although the observations were carried out with the entire LOFAR array (Dutch array and international stations), we only use the data from the Dutch array here to image the large-scale structure. We did not use this international stations data for the high resolution image because the target was too far from the phase centre for high fidelity high resolution imaging. 
The data were processed using the direction dependent self-calibration pipeline, DDF-pipeline\footnote{\url{https://github.com/mhardcastle/ddf-pipeline}}, described in detail in \citet{Shimwell2019b,Tasse2020a}. The flux scale is consistent with LoTSS-DR2. LoTSS-DR2 is scaled to \citet{Roger1973} flux scale (consistent with \citealt{Scaife2012}) through statistically aligning to the 6C survey assuming a global NVSS/6C spectral index. \par
To improve image quality, the target was extracted from the self-calibrated data after subtracting from the uv-data, all the sources located in the field of view other than 3C\,293 (and a few other sources nearby). Additional self-calibration and imaging loops were then performed on the extracted data set \citep{VanWeeren2020} using WSClean for imaging and NDPPP for calibration. The final image has a beam size of 10.5\arcsec$\times$7\arcsec and has an off-source RMS of 0.3\mjybeam. The RMS increases to 2 \mjybeam close to the target and this local RMS noise has been used hereafter. This new low frequency radio map is shown in Figure~\ref{fig:hba_low}. We measure an integrated flux density of 15.19$\pm$1.52\,Jy within 3$\sigma_\mathrm{RMS}$ contours and use a 10\% flux density scale error. This value is in agreement with the integrated flux density of 15.0$\pm$1.5\,Jy from TIFR GMRT Sky Survey (TGSS) at 154\,MHz \citep{Intema2017}. \par
From Figure~\ref{fig:hba_low}, it can be seen that the central region shows an elongation in the north-south direction. This elongation shows up after the extraction process on the LoTSS-DR2 data set and is not real. The rest of the structure in the map matches very well with higher frequency maps and therefore does not seem to be affected by the extraction.

\subsection{Archival data}

\subsubsection{Low resolution VLA 1.36\,GHz and 4.85\,GHz}
We reprocessed archival VLA 1.36\,GHz and 4.85\,GHz data for the target. 1.36\,GHz observations were carried out with the VLA in B configuration in November 1999 (Project code - GP022). The target was observed for $\sim$11 hours with a 25\,MHz bandwidth. 3C~48 and OQ~208 were the flux and phase calibrators respectively. 4.85\,GHz observations of 3C\,293 were carried out on November 1986 (Project code AB412) and October 1984 (AR115) in C and D configuration respectively. 3C~286 was used as the flux and phase calibrator for the AB412 data set. 3C~286 and OQ~208 were used as the flux and phase calibrator respectively for the AR0115 data sets. These were combined after cross-calibration was done individually on each data set to include both long baseline for a high resolution and short baselines to recover large-scale emission from the source.\par
All data sets were reduced up till cross-calibration in Astronomical Image Processing System (AIPS, \citealt{Greisen2003}). The data were manually flagged and the flux scale was set according to \citet{Perley2013} as it is consistent with the \citet{Scaife2012} scale at low frequencies. The self-calibration and imaging were performed in Common Astronomy Software Applications (CASA, \citealt{McMullin2007}) using the standard procedure\footnote{\url{https://casaguides.nrao.edu/index.php/VLA_Self-calibration_Tutorial-CASA5.7.0}} and the final images were obtained using Briggs weighting with a robust parameter of $-$0.5. The RMS noise is 0.08 \mjybeam and 0.22 \mjybeam for the 1.36\,GHz and 4.85\,GHz image respectively. 

\subsubsection{High resolution MERLIN} 
We have used the high resolution MERLIN image of the centre at 1.36\,GHz made by \citet{Beswick2002} (data reduction procedure described therein), which has a resolution of 0.23\arcsec$\times$0.20\arcsec. The integrated flux density of the central region estimated from the MERLIN image was 4.1\,Jy, which is significantly higher than the integrated flux density of the central region from the low resolution VLA 1.36\,GHz image ($\sim$3.75\,Jy). To align the fluxes with previous measurements, we have used the VLA 1.36\,GHz image at 1\arcsec resolution from \citet{Mahony2013The293}. We estimated the integrated flux density in the two images using pyBDSF \citep{Mohan2015} and then scaled the MERLIN image to the 1\arcsec image. We used a flux scale error of 5\%. The final integrated flux density of the MERLIN image was 3.72$\pm$0.19\,Jy with a local RMS noise of 1.2 \mjybeam.

\subsubsection{High resolution VLA 4.85\,GHz and 8.45\,GHz }
We have reprocessed VLA 4.85\,GHz archival data to make a high resolution image of the central region at 4.8\,GHz. 3C\,293 was observed with the VLA in December 2000 (AT0249) and August 2002 (AH0766) in A and B configuration respectively. 3C~286 and OQ~208 were used as flux and phase calibrators respectively for both data sets. \par
VLA 8.45\,GHz data for 3C\,293 from observations on September 1991 (A configuration, Project code - AO0105) and December 1995 (B configuration, Project code - AK0403) was also reprocessed. 3C~286 was used as the flux calibrators for both data sets while 1144+402 and 1607+268 were used as the phase calibrators for the September 1991 and December 1995 observation respectively. \par
The flagging and cross-calibration for all archival VLA data sets were performed in AIPS and the flux scale was set according to \citet{Perley2013}. The cross-calibrated data sets were then combined to obtain a good uv-coverage and taken to CASA for self-calibration and imaging using Briggs weighting with a robust parameter of -0.5. The image statistics are summarised in Table~\ref{image_data}. The flux scale errors used are 5\% for the VLA 4.85\,GHz and 8.45\,GHz data.

\begin{figure}
\includegraphics[width=\columnwidth]{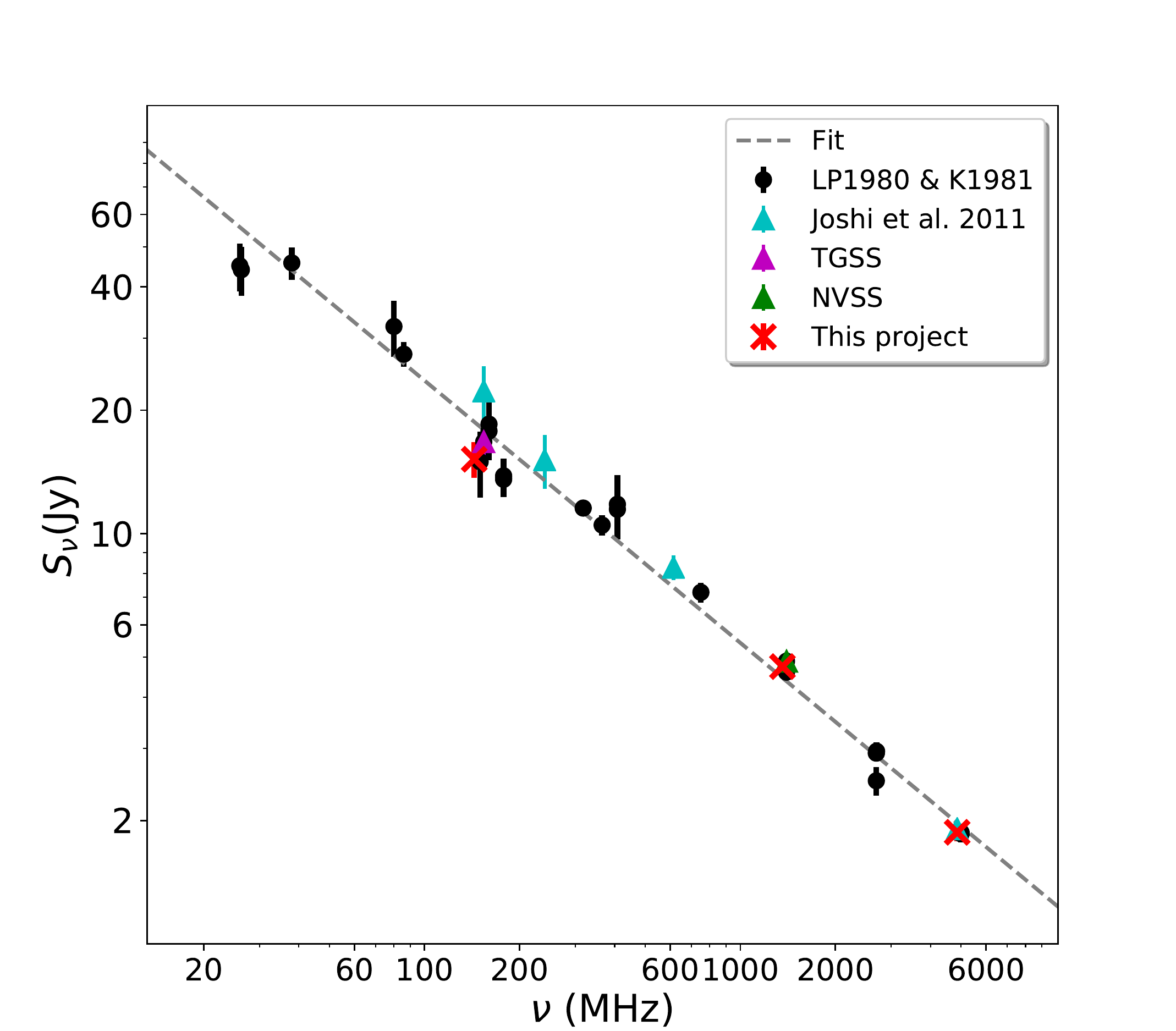}
\caption{Integrated flux spectrum for 3C\,293 from our images and literature. The red crosses are the integrated flux densities measured within the 3\,$\sigma$ contours in our images, where $\sigma$ is the local RMS noise. For the LOFAR HBA point at 144\,MHz, only Dutch array data was used. Black circles are values from \citealt{Laing1980} (LP1980) and \citealt{Kuehr1981} (K1981). Cyan, purple and green triangles are values from the \citet{Joshi2011}, TGSS \citep{Intema2017} and NVSS \citep{Condon1998} respectively. Flux densities from TGSS and this project were set to the \citet{Baars1977} flux scale for consistency. The dashed line shows the straight line fit using the values from literature (including TGSS and NVSS) and has a slope of $-$0.64$\pm$0.01.}
\label{int_spec}%
\end{figure}
\begin{figure*}
        \centering
        \begin{subfigure}[b]{0.505\textwidth}
            \centering
            \includegraphics[width=\textwidth]{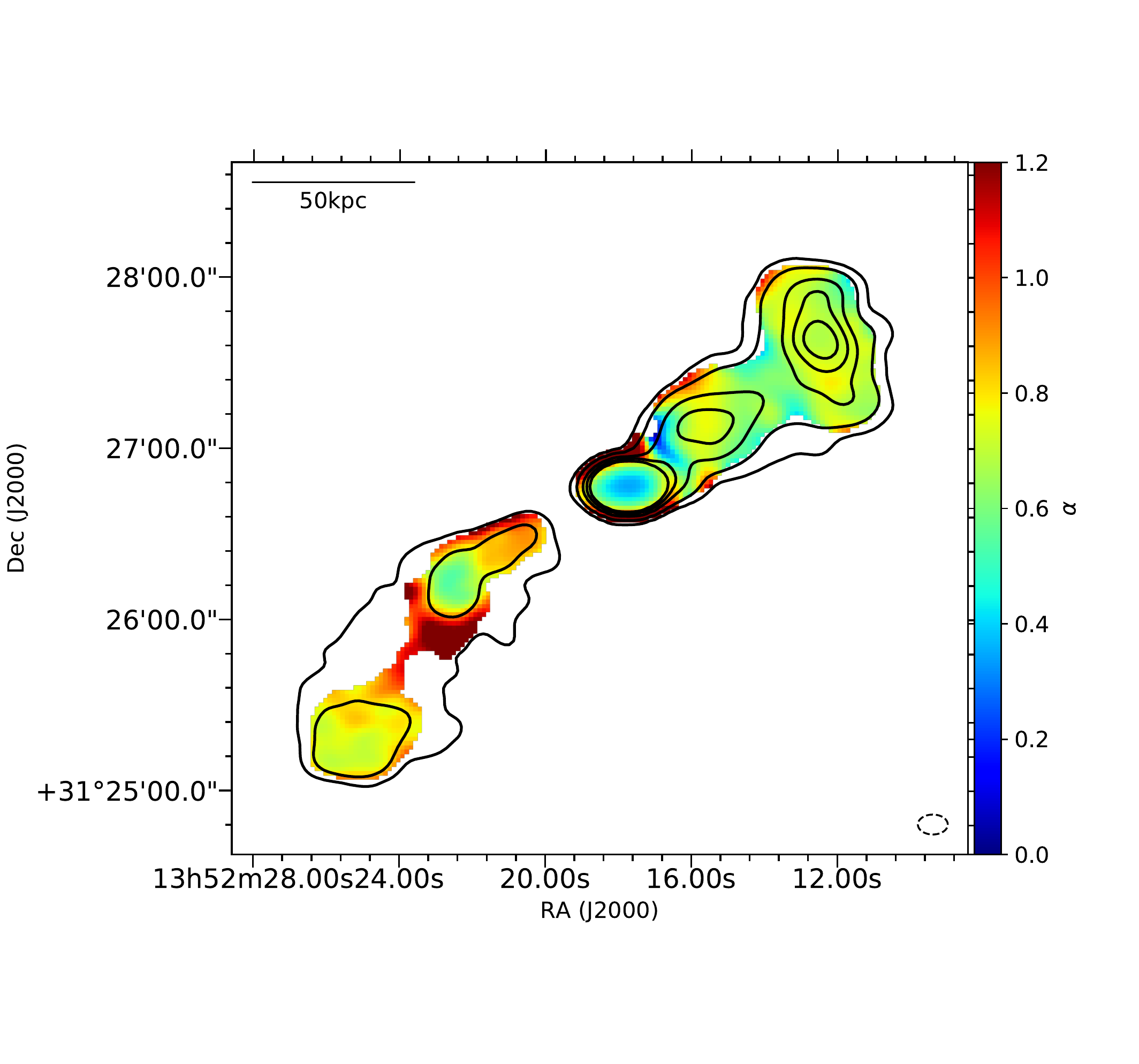}
            \caption[1]%
            {{\small Spectral index map $\alpha^\mathrm{144}_\mathrm{1360}$}}    
            \label{spix_144_1400}
        \end{subfigure}
        \hfill
        \begin{subfigure}[b]{0.485\textwidth}  
            \centering 
            \includegraphics[width=\textwidth]{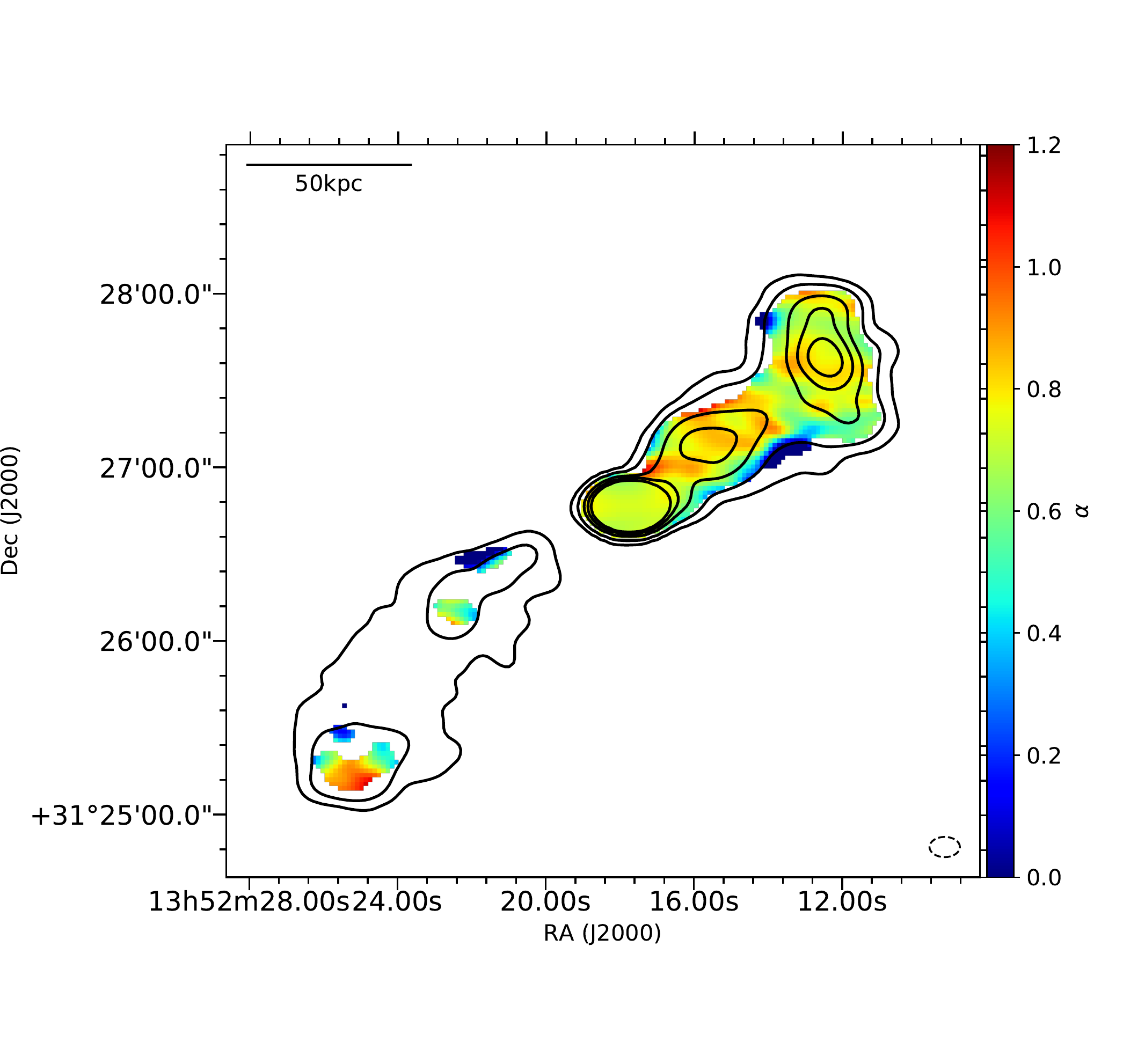}
            \caption[2]%
            {{\small Spectral index map for $\alpha^\mathrm{1360}_\mathrm{4850}$}}  
            \label{spix_1400_4850}
        \end{subfigure}
        \caption[The spectral index maps from 144-4850\,MHz.]
        {\small (a) The spectral index map from 144-1360\,MHz and (b) 1360-4850\,MHz for the large-scale lobes emission of 3C\,293. All the maps are overlaid with contours of the 1360\,MHz image at 10.5\arcsec$\times$7\arcsec resolution with the contour levels at (5, 30, 150, 350, 700)$\times\sigma_\mathrm{RMS}$ where $\sigma_\mathrm{RMS}$=0.09 \mjybeam is the local RMS noise. A flatter spectrum with $\alpha^\mathrm{144}_\mathrm{1360}$ < 0.5 can be seen in the centre of the 144-1360\,MHz map. The 1360-4850\,MHz spectral index map does not include most of the southern lobe, since very little emission from the southern lobe is recovered at 4850\,MHz.} 
        \label{spix}
\end{figure*}

\subsection{Flux density scale}
The accuracy of the flux scale is a key requirement for spectral index studies. To first confirm the flux scales of all our low resolution images, we have plotted the integrated flux densities of the entire source (centre + outer lobes) from our images along with values from literature \citep{Laing1980,Kuehr1981} in Figure~\ref{int_spec}. Flux densities from the TGSS at 154\,MHz \citep{Intema2017} and NVSS at 1400\,MHz \citep{Condon1998} were also plotted alongside. The flux densities for our maps and a best fit line to the literature values are also shown. Although our integrated flux density at 144\,MHz is in good agreement with the TGSS, as mentioned before, and other values from literature, it is significantly different from \citet{Joshi2011}, who estimate the integrated flux density to be 22.3$\pm$3.4\,Jy at 154\,MHz. The difference between the flux density estimated from the best fit (dashed line in Figure~\ref{int_spec}) and the measured flux density ($\Delta$S) is lower for our 144\,MHz value ($\Delta$S$\approx$3.2\,Jy) than the 154\,MHz \citet{Joshi2011} value ($\Delta$S$\approx$4.4\,Jy). Due to the agreement of our flux densities with most of the literature, we consider our flux scale correct despite the discrepancy with \citet{Joshi2011}.\par
As a sanity check, we have compared the integrated flux densities of the target from the high resolution images (see Table~\ref{image_data}) with the integrated flux density of region C (see Table~\ref{region_summary}) from the low resolution images. We find that the flux scale is in agreement between the two images, which is in turn in agreement with literature.\par
Throughout the paper, errors in flux densities have been calculated using a quadrature combination of the noise errors and the flux scale errors. The noise error depends on the size of the integration area $A_\mathrm{int}$ (integration area in units of beam solid angle) and the RMS noise $\sigma_\mathrm{RMS}$ in the map as $\Delta S_\mathrm{n}=\sigma_\mathrm{RMS}\times\sqrt{A_\mathrm{int}}$ \citep{Klein2003}.
We use a flux scale error of 10\% for LOFAR HBA, and 5\% for VLA 1.36\,GHz, 4.85\,GHz and 8.45\,GHz data \citep{Scaife2012,Perley2013}. 

\subsection{Spectral index maps}
To spatially resolve the spectral properties, we have constructed spectral index maps using the low resolution images. Spectral index maps require images at the same resolution and range of uv-spacings, therefore all the images were smoothed to have a common resolution. Since an interferometer with minimum baseline D$_{min}$ is sensitive to a largest angular scale of 0.6 $\lambda$/D$_{min}$ \citep{Tamhane2015}, we ensured that all the data we used was sensitive to angular scales $\gtrsim$250\arcsec, which is the angular size of 3C\,293.

We first smoothed the VLA 1.36\,GHz and 4.85\,GHz images to a resolution of 10.5\arcsec$\times$7\arcsec (BPA=90$\degree$), which is the LOFAR resolution and corresponds to a physical scale of $\sim$9$\times$6\,kpc at the source redshift. \par
Phase calibration can cause position offsets between images of different frequencies. Such offsets can lead to systematic artefacts in the spectral index maps and therefore it is necessary to align the images and correct for such offsets. Ideally, point sources near the target are suited for this purpose, but since we do not have them in our field at all frequencies, we used the position of the peak flux density of the central region as reference. We fitted a 2D Gaussian on the central region to derive the pixel position coordinates of the peak flux density. We then used the coordinates of one image as reference and aligned all the others to it using the tasks IMHEAD and IMREGRID in CASA. After this, the positions matched with an accuracy of $\leq$0.01 pixels, which is sufficient for our analysis. Spectral index maps were then made using the task IMMATH in CASA using only the emission within the 5\,$\sigma$ contours. The 144-1360\,MHz and 1360-4850\,MHz maps are shown in Figure~\ref{spix}. The 144-1360\,MHz map shows steep index regions in the north and south edges of the centre, which are due to the artefact in the 144\,MHz map, as discussed in Sect. 2.1.2.
\par

\begin{figure}
   \includegraphics[width=\columnwidth]{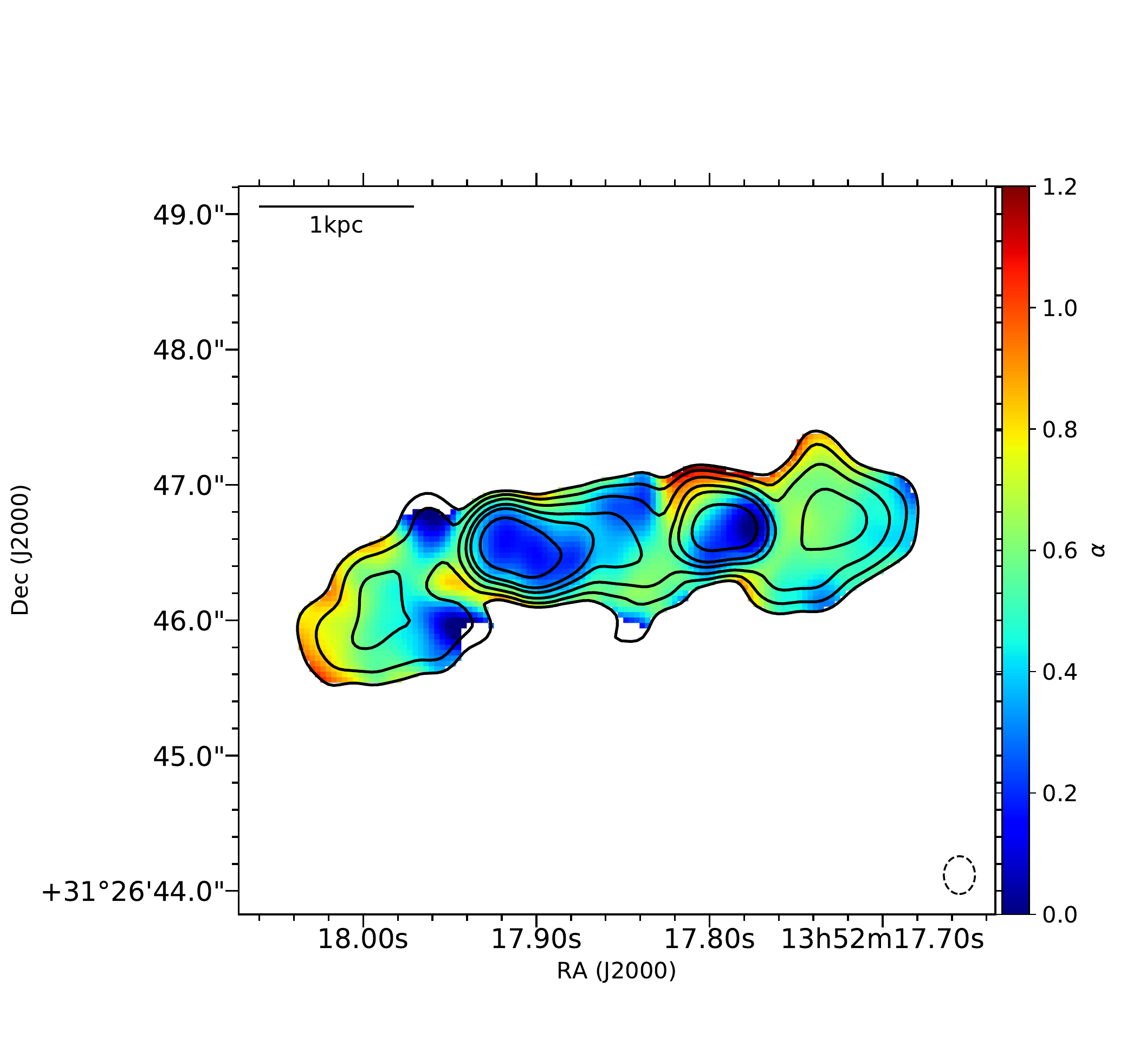}
   \caption{Spectral index map of the high resolution central region of 3C\,293 from 144-1360\,MHz at 0.28\arcsec$\times$0.23\arcsec resolution. The map is overlaid with 1360\,MHz contours with levels at: (5, 10, 20, 40, 65, 150)$\times\sigma_\mathrm{RMS}$ where $\sigma_\mathrm{RMS}$ = 1.4 \mjybeam is the local RMS noise in the smoothed MERLIN image. The inner lobes clearly show a distinct spectral index population from the diffuse emission around them.}
    \label{spix_highres}%
\end{figure}

The same procedure was followed to make spectral index maps using the high resolution images of the central region. We used the pixel position of the peak flux density of the eastern inner lobe for aligning the images. The 144-1360\,MHz spectral index map is shown in Figure~\ref{spix_highres}.  
At higher frequencies, the quality of our higher frequency spectral index maps for the centre was not good enough to allow us to perform a pixel by pixel analysis. Therefore, to probe the spectral properties, we have extracted integrated flux densities in regions across the centre, as shown in Figure~\ref{fig:region}. These flux densities were then used to calculate the spectral index in the regions. Regions E1, W1 and E2, W2 are on the inner lobes and diffuse emission respectively and O1 and O2 are on the outer north-western lobe. Comparing these regions to the high resolution image in \citet{Beswick2004}, the core would lie in the W1 region. These regions and their sizes can be seen in Figure~\ref{fig:region} and the extracted flux densities and spectral indices in Table~\ref{region_summary}. An integrated spectrum from 144\,MHz to 4850\,MHz comparing the central and outer lobe regions is shown in Figure~\ref{region_spix}.\par

Throughout the paper, errors in the spectral indices are calculated as :
\begin{ceqn}
\begin{align}
      \Delta\alpha =
          \frac{1}{\ln(\frac{\nu_\mathrm{1}}{\nu_\mathrm{2}})}\sqrt{\left(\frac{\Delta S_\mathrm{1}}{S_\mathrm{1}}\right)^\mathrm{2}+\left(\frac{\Delta S_\mathrm{2}}{S_\mathrm{2}}\right)^\mathrm{2}},
\end{align}
\end{ceqn}

\noindent where $\Delta S_\mathrm{1}$ and $\Delta S_\mathrm{2}$ are errors in the flux densities which include statistical errors in the measurements as well as uncertainties in the overall scale. 

\section{Results}
In this section, we first describe the morphology of the large and small-scale emission in 3C\,293. Then we estimate the magnetic field values for the source which are needed for the spectral ageing analysis. Then we describe the spectral index properties and the modelling of the spectrum.
\subsection{Morphology}
The large-scale morphology of the outer lobes and the small-scale morphology of the centre are discussed in the following sections.
\subsubsection{Outer lobes}
The large structure of 3C\,293 is shown in Figure~\ref{lofar_hba} at 144\,MHz. The large-scale morphology is in agreement with what has been seen before at higher frequencies \citep{Bridle1981,Joshi2011}. The outer lobe emission in Figure~\ref{fig:hba_low} is $\sim$250$\arcsec$ in total extent which corresponds to a physical size of $\sim$220\,kpc at the source redshift (measured using 3\,$\sigma$ contours as a boundary).
The morphology is made up of a bright central region and two outer lobes, one in the north-west ($\sim$84\,kpc) and another in the south-east ($\sim$107\,kpc) direction. Despite increased sensitivity we do not obviously detect any new features in the low resolution LOFAR map. The bright emission at the end of the north-western lobe, covered by O2, extends in a direction perpendicular to the axis of the lobe. This bright region was suggested to be a hotspot before (for example \citealt{Joshi2011,Lanz2015}), although as discussed in Sec 3.2, the spectral properties do not confirm this. Using our 4.5\arcsec$\times$4.1\arcsec resolution 1.4\,GHz image, we also measure a physical size of $\sim$20\,kpc for this region, which is larger than the typical sizes of $\lesssim$10\,kpc for hotspots (for example \citealt{Jeyakumar2000}). These properties suggest that the large-scale radio emission in 3C\,293 is not that of a typical FRII radio galaxy.\par
\begin{figure}
   \includegraphics[width=\columnwidth]{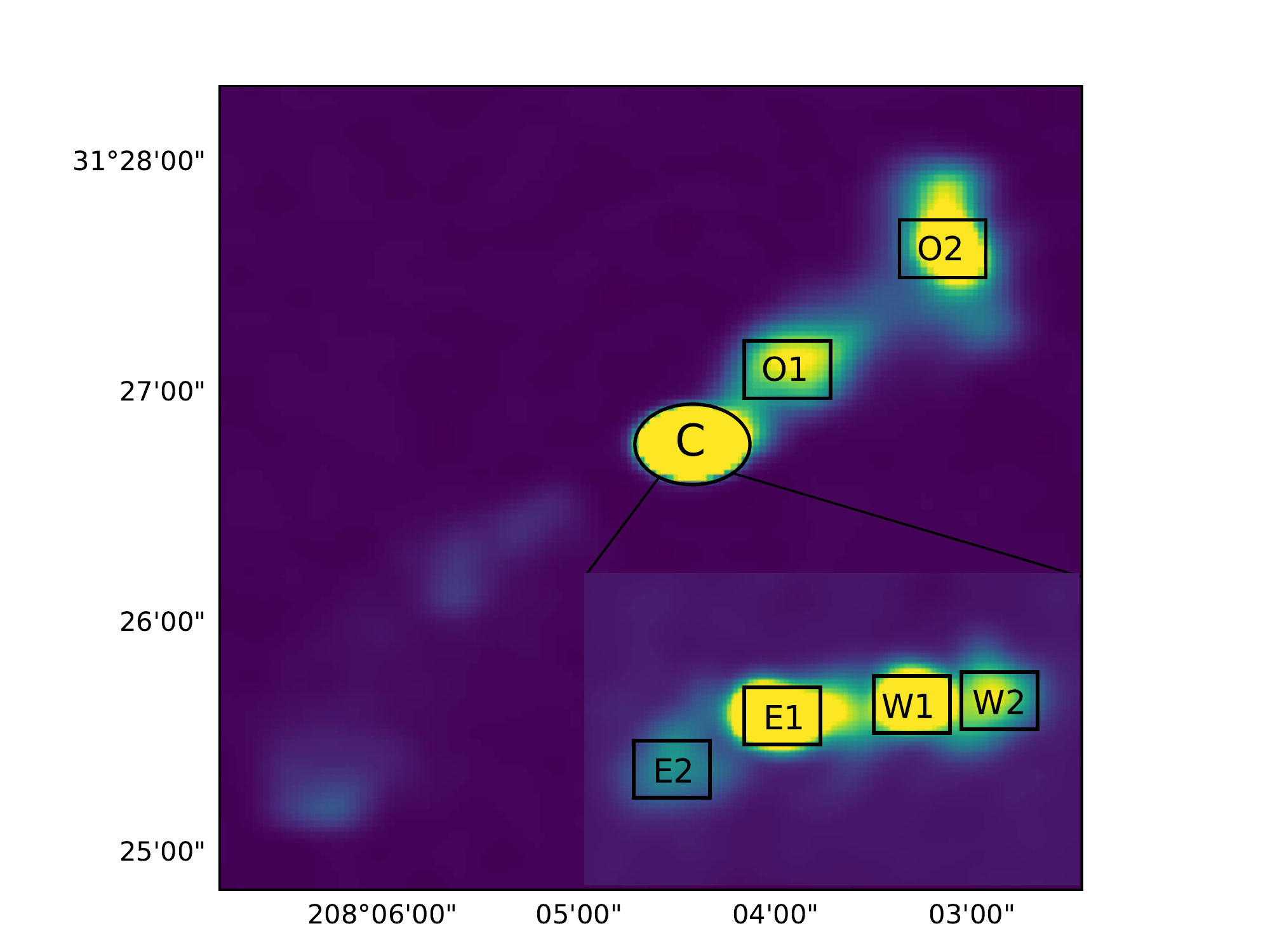}
   \caption{Regions used for extracting the fluxes in the low and high resolution images. Regions O1 and O2 of size 20\arcsec$\times$15\arcsec and C of size 40\arcsec$\times$25\arcsec were used to extract the fluxes after smoothing all images to the low resolution of 10.5\arcsec$\times$7.0\arcsec. Regions E1, E2, W1 and W2 of size 0.8\arcsec$\times$0.6\arcsec were used to extract the fluxes from the high resolution images at a resolution of 0.37\arcsec$\times$0.36\arcsec. The continuum maps are at 1360\,MHz.}
    \label{fig:region}%
\end{figure}
The south-eastern lobe also shows two regions of bright emission, although its total emission is fainter than its counterpart in the north. This asymmetry in intensity of the two lobes makes it very hard to image the south-eastern lobe and we do not have enough sensitivity at VLA 4.85\,GHz to fully recover this emission. By looking at the contours, we can see a difference in the morphology of the emission bridge that connects the two bright emission regions at the ends of the lobe at 144\,MHz (Figure~\ref{fig:hba_low}) and 1360\,MHz (see contours in Figure~\ref{spix_144_1400}). The low significance of this feature and the uncertainty over its origin mean that at this moment, we are hesitant to conclude whether this difference in morphology is real or introduced during the self-calibration and imaging.

 \begin{table*}[h]
   \centering
  \begin{threeparttable}
       \caption[]{Flux densities and spectral indices for regions.}
         \label{region_summary}
     \begin{tabular}{cccccccc}
            \hline
            \hline
            \noalign{\smallskip}
            
            Region & $S_\mathrm{144\,MHz}$  & $S_\mathrm{1360\,MHz}$ & $S_\mathrm{4850\,MHz}$ & $S_\mathrm{8450\,MHz}$ & $\alpha^\mathrm{144}_\mathrm{1360}$& $\alpha^\mathrm{1360}_\mathrm{4850}$ & $\alpha^\mathrm{4850}_\mathrm{8450}$ \\
             & (Jy) & (Jy) & (Jy) & (Jy) & & &\\ 
            \noalign{\smallskip}
            \hline
            \hline
            \noalign{\smallskip}
    \vspace*{5px}
    E1  & 2.37$\pm$0.24 & 1.27$\pm$0.06 & 0.65$\pm$0.03 & 0.46$\pm$0.02 & 0.28 & 0.52 & 0.63 \\
    \vspace*{5px}
    W1  & 2.32$\pm$0.23 & 0.97$\pm$0.05 & 0.33$\pm$0.02 & 0.21$\pm$0.01 & 0.39 & 0.84 & 0.80\\
    \vspace*{5px}
    E2  & 0.63$\pm$0.06 & 0.16$\pm$0.01 & 0.05$\pm$0.002 & 0.03$\pm$0.002 & 0.61 & 0.95 & 0.72 \\
    \vspace*{5px}
    W2  & 1.04$\pm$0.10 & 0.29$\pm$0.01 & 0.08$\pm$0.004 & 0.05$\pm$0.002 & 0.57 & 1.02 & 0.95\\
    \vspace*{5px}
    O1 & 0.64$\pm$0.07 & 0.13$\pm$0.01 & 0.05$\pm$0.003 & - & 0.72 & 0.76 & -\\
    \vspace*{5px}
    O2 & 1.09$\pm$0.11 & 0.22$\pm$0.01 & 0.08$\pm$0.004 & - & 0.71 & 0.78 & -\\
    \vspace*{5px}
    C & 10.70$\pm$1.07 & 3.84$\pm$0.19 & 1.53$\pm$0.08 & - & 0.46 & 0.72 & -\\
            \noalign{\smallskip}
            \hline
            \hline
     \end{tabular}      
    
      \small{Region column lists the regions as shown in Figure~\ref{fig:region}. The high resolution image (0.37\arcsec$\times$0.36\arcsec) was used for the E1, E2, W1 and W2 and the low resolution image was used for O1, O2 and C. $S_\mathrm{144\,MHz}$, $S_\mathrm{1360\,MHz}$, $S_\mathrm{4850\,MHz}$ and $S_\mathrm{8450\,MHz}$ columns list the integrated flux densities, extracted from the regions at different frequencies.  $\alpha^\mathrm{144}_\mathrm{1360}$, $\alpha^\mathrm{1360}_\mathrm{4850}$ and $\alpha^\mathrm{4850}_\mathrm{8450}$ columns list the spectral indices. The 1\,$\sigma$ errors in the spectral indices are $\pm$0.05, $\pm$0.06 and $\pm$0.13 for  $\alpha^\mathrm{144}_\mathrm{1360}$, $\alpha^\mathrm{1360}_\mathrm{4850}$ and $\alpha^\mathrm{4850}_\mathrm{8450}$ respectively.}
     \end{threeparttable} 
   \end{table*}

\subsubsection{Central region} 
In Figure~\ref{fig:hba_high}, we show the high resolution map with LOFAR international stations at 144\,MHz. This is the first time a resolved map of the central region has been made at such low frequencies. The total extent of the source is $\sim$4.5\,kpc in physical size made up of the two compact bright emission regions (referred to as the inner lobes henceforth), with lower brightness diffuse emission on either side.

We do not detect any obvious new features in our low frequency map compared to high frequencies. The inner lobes have a physical size of $\sim$2\,kpc in projection, with the core of the AGN located in the western inner lobe in region W1 of Figure~\ref{fig:region} \citep{Beswick2004}.\par

\begin{figure}
   \includegraphics[width=\columnwidth]{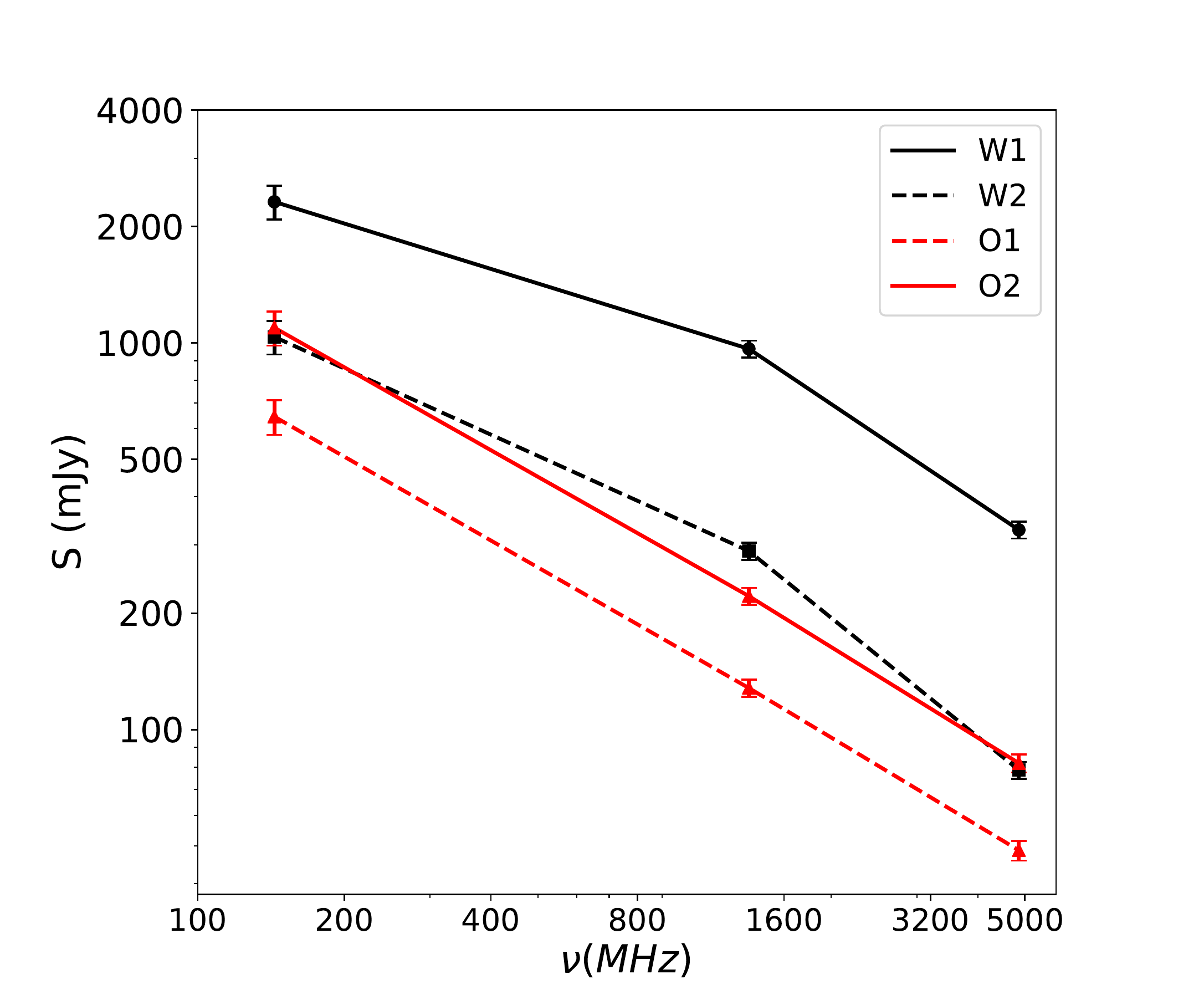}
   \caption{Integrated spectra for the regions in the centre along the western lobe and outer lobe. Black lines show the spectra for central regions and red lines for outer lobe regions.}
    \label{region_spix}%
\end{figure}

\subsection{Spectral properties}

\subsubsection{Outer lobes}
To perform a resolved study of the spectral index in the outer lobes, we have constructed the spectral index maps shown in Figure~\ref{spix} in the range 144-4850\,MHz. The $\alpha^\mathrm{144}_\mathrm{1360}$ map (Figure~\ref{spix_144_1400}) shows a typical spectral index of 0.6-0.7 ($\pm$0.07) in the north-western lobe. In the $\alpha^\mathrm{1360}_\mathrm{4850}$ map (Figure~\ref{spix_1400_4850}) this steepens to 0.7-0.8 ($\pm$0.06). As mentioned in Sect 2.5, we have also used regions O1 and O2 (see Figure~\ref{fig:region}) to calculate spectral indices. We find that the spectrum of the north-western lobe steepens from $\alpha^\mathrm{144}_\mathrm{1360}$=0.72$\pm$0.05 to $\alpha^\mathrm{1360}_\mathrm{4850}$ = 0.76$\pm$0.06 for O1 and from $\alpha^\mathrm{144}_\mathrm{1360}$=0.71$\pm$0.05 to $\alpha^\mathrm{1360}_\mathrm{4850}$ = 0.78$\pm$0.06 for O2. The integrated spectrum of these regions can be seen in Figure~\ref{region_spix}.

The first property to note here is the small curvature of the spectral index throughout the $\sim$90\,kpc north-western lobe from 144-1360\,MHz to 1360-4850\,MHz, of $\Delta\alpha\leq0.2$. There is also no sign of an ultra-steep spectrum ($\alpha^\mathrm{144}_\mathrm{1360}$ or $\alpha^\mathrm{1360}_\mathrm{4850}$ > 1.2) anywhere in the north-western lobe.
The second property is the lack of a spectral gradient throughout this lobe, with increasing distance from the centre, as can be seen in Figure~\ref{spix}. The spectral index appears to have a homogeneous distribution, at our resolution, which corresponds to a physical scale of $\sim$10\,kpc. Another interesting property to note is the spectral index at the bright region at the end of the lobe (O2), which has been identified as a hotspot by previous studies. The spectral index of this emission does not show any flattening as would be expected from a hotspot. Despite the difference in flux density noted in Section 2.3, our spectral index results are consistent with \cite{Joshi2011}.\par

In the south-eastern lobe, $\alpha^\mathrm{144}_\mathrm{1360}$ is different from the north-western lobe. The outer edge has a spectral index of 0.7$\pm$0.05 which steepens to 0.8-0.9 ($\pm$0.05) moving towards the core. We see a region of very steep (dark red) spectral index ($\alpha^\mathrm{144}_\mathrm{1360}$ > 1) just before the intermediate knot. This knot has a flatter spectral index of 0.6$\pm$0.05 followed by a region of 0.9$\pm$0.05 index moving towards the core. This spectral index distribution with multiple regions of flat and steep index is similar to that seen in FRIIs with episodic activity.  The dark red spectral index region in this lobe is a result of the difference in spatial distribution of the flux density at 144\,MHz and 1360\,MHz and we do not have confidence in this feature, as described in Sect. 3.1.1. Furthermore, since we do not recover most of the emission in this lobe at 4850\,MHz, we do not have confidence in the spectral index in Figure~\ref{spix_1400_4850} and cannot perform spectral age modelling for it but discuss the spectral index from 144-1360\,MHz for some regions in this lobe in Section 4.2. \par

\subsubsection{Central region}
The central region in the spectral index maps in Figure~\ref{spix} shows an index of $\alpha^\mathrm{144}_\mathrm{1360}$=0.4-0.5 ($\pm$0.05). This steepens to $\alpha^\mathrm{1360}_\mathrm{4850}$=0.7-0.8 ($\pm$0.06). To investigate the spectrum in this region in more detail, we have used the higher resolution images.  The resulting spectrum is shown in Figure~\ref{region_spix}.\par 

As can be seen in Table~\ref{region_summary}, $\alpha^\mathrm{144}_\mathrm{1360}$ is 0.28$\pm$0.05 and 0.39$\pm$0.05 for the inner lobes regions E1 and W1 respectively, and $\alpha^\mathrm{1360}_\mathrm{4850}$ steepens to 0.52$\pm$0.06 and 0.84$\pm$0.06 respectively. These spectral index values are consistent with the spectral index value obtained by integrating over the entire central region C. The low frequency index of regions E1 and W1 is likely affected by absorption as it is well below the theoretical limit for injection index, which typically is 0.5. For the diffuse emission regions E2 and W2, $\alpha^\mathrm{144}_\mathrm{1360}$ is 0.61$\pm$0.05 and 0.57$\pm$0.05, respectively, while $\alpha^\mathrm{1360}_\mathrm{4850}$ is 0.95$\pm$0.06 and 1.02$\pm$0.06, respectively. Therefore, the central regions show a sharp break in their spectrum around 1360\,MHz, and overall, the spectrum is steeper for the western regions in comparison to the eastern regions. The diffuse emission regions also show a steeper spectra than the inner lobe regions. 
A similar distinction in the spectra was also seen between 1.7-8.4\,GHz by \citet{Akujor1996b}. This difference in the spectral index suggests a difference in the nature of these components and we discuss the absorption in section 3.4.2.\par
\subsection{Magnetic field} 
The magnetic field is a crucial input to spectral ageing models, as the strength of the magnetic field affects how much the spectrum steepens over a period of time. Indeed, estimating magnetic fields for radio galaxies is difficult and usually, a simplifying assumption of equipartition between relativistic particles and the magnetic field is used (for example \citealt{Jamrozy2007,Konar2012,Nandi2010,Nandi2019,Sebastian2018}). A detailed derivation for the equipartition magnetic field is given in \citet{Worrall2004}. X-ray emission from the outer lobes is used for a more direct probe for the magnetic field strength, since it is understood to originate from the inverse-Compton scattering between the relativistic electrons and the CMB photons (for example \citealt{Feigelson1995,Isobe2002,Hardcastle2002,Mingo2017}). In the last two decades, studies using IC-CMB X-ray emission have found that for FRII galaxies, magnetic field strengths are lower than the equipartition values by a factor of 2-3 \citep{Croston2005,Ineson2017,Turner2017}. However, the X-ray studies of some radio galaxies have also found magnetic field strengths close to the equipartition value (for example \citep{Croston2005,Konar2009}).
Since 3C\,293 is not a typical FRII galaxy and X-ray studies of 3C\,293 have only detected emission from small regions of the outer lobes \citep{Lanz2015}, which is not enough for this purpose, we have used the equipartition assumption. We acknowledge that the actual magnetic field in the outer lobes of 3C\,293 might be sub-equipartition and discuss its effect on the derived spectral ages in Section 4.2. 
We have used the pySYNCH code\footnote{\url{https://github.com/mhardcastle/pysynch}}, which is a python version of the SYNCH code \citep{Hardcastle1997}, to estimate the magnetic field strength in our synchrotron source (also see \citealt{Mahatma2020}). \par
For the outer north-western lobe, the integrated flux densities were used to fit a synchrotron spectrum. The lobe was assumed to be of cylindrical geometry with a length of 90\arcsec and radius of 20\arcsec (size measured using 5$\sigma$ contours for reference). For the particle energy distribution, a power law of the form $N(\gamma) \propto \gamma^\mathrm{-p}$ was used where $\gamma$ is the Lorentz factor in the range $\gamma_\mathrm{min}$ = 10 to $\gamma_\mathrm{max}$ = 10$^\mathrm{6}$ and $p$ is the particle index and is related to the injection index in the synchrotron spectrum as $p$ = 2$\alpha_\mathrm{inj}$+1. We used Broadband Radio Astronomy Tools (BRATS\footnote{\url{http://www.askanastronomer.co.uk/brats/}}; \citealt{Harwood2013,Harwood2015}) package to find the best fit injection index over the north-western lobe, which gave a best fit injection index of $\alpha_\mathrm{inj}$ = 0.61$\pm$0.01 (see Figure~\ref{injectionmap} and Sect. 3.4). This gives an equipartition magnetic field of $B_\mathrm{NW}$ = 5.9$^{+0.14}_{-0.13}$ $\mu$G. This value is similar to the magnetic field strength of 5\,$\mu$G estimated by \citet{Machalski2016a}. \citealt{Joshi2011} estimate a magnetic field of 11.2 $\mu$G, which is $\sim$2 times higher than our estimate. This discrepancy is due to the difference in flux density discussed in Section 2.3. The magnetic field estimated for our data here is used as an input for the spectral ageing models discussed in Sect. 3.4.\par
We also estimated the equipartition magnetic field for the central regions shown in Figure~\ref{fig:region} using a cylindrical geometry with a length of 0.8\arcsec and a radius of 0.3\arcsec. An aged synchrotron spectrum was used to estimate the equipartition field instead of a powerlaw because it provided a better match to the curved spectra seen in these regions.
For the inner lobe regions, the 144\,MHz data point was not used because it is affected by absorption, which makes it very hard to estimate the injection index. We used an injection index of $\alpha_\mathrm{inj}$ = 0.5, which is the theoretical limit for injection indices found for plasma in radio galaxies because absorption in the inner lobes (see Sect. 3.5.1 and 4.1.1) erases any information about injection. 
For regions E1 and W1, $B_\mathrm{E1}=$ 237$^{+13}_{-9}$ $\mathrm{\mu}$G and $B_\mathrm{W1}=$ 218$^{+17}_{-14}$ $\mu$G were obtained respectively. For the diffuse emission regions, the low frequency index was used as the injection index, however it is possible that this value is affected by absorption and the actual injection index is steeper. For E2 and W2, $B_\mathrm{E2}=$ 158$^{+15}_{-13}$ $\mu$G and $B_\mathrm{W2}=$ 176$^{+16}_{-14}$ $\mu$G were obtained respectively. To check the robustness of our estimates, we repeated the calculations using regions of various sizes and obtained field strengths that are in agreement with each other.

\subsection{Modelling of spectra}
\subsubsection{Spectral age modelling} 
\label{specage}
\begin{figure}
   \includegraphics[width=\columnwidth]{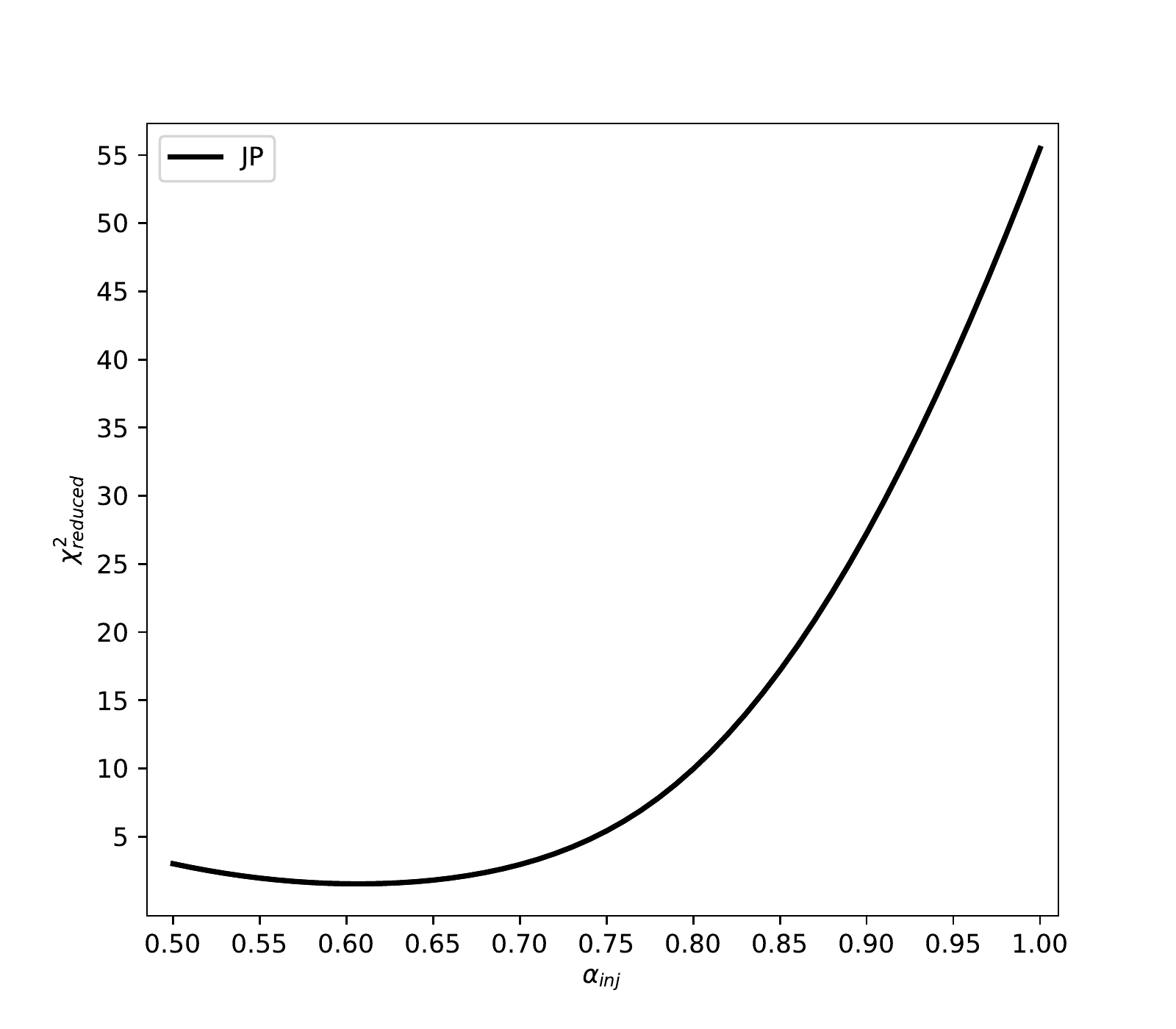}
   \caption{Reduced $\chi^\mathrm{2}$ values for the varying injection index values for the north-western lobe of 3C\,293, averaged over the regions. The data points are taken at an interval of 0.01 between 0.5 and 1.0, and the minimum occurs at 0.61 for both JP and Tribble models. The reduced $\chi^\mathrm{2}$ values for the two models are very similar, therefore only the JP model values are shown here. }
    \label{injectionmap}
\end{figure}

\begin{figure*}
\begin{subfigure}{.34\textwidth}
  \centering
  \includegraphics[width=1\linewidth]{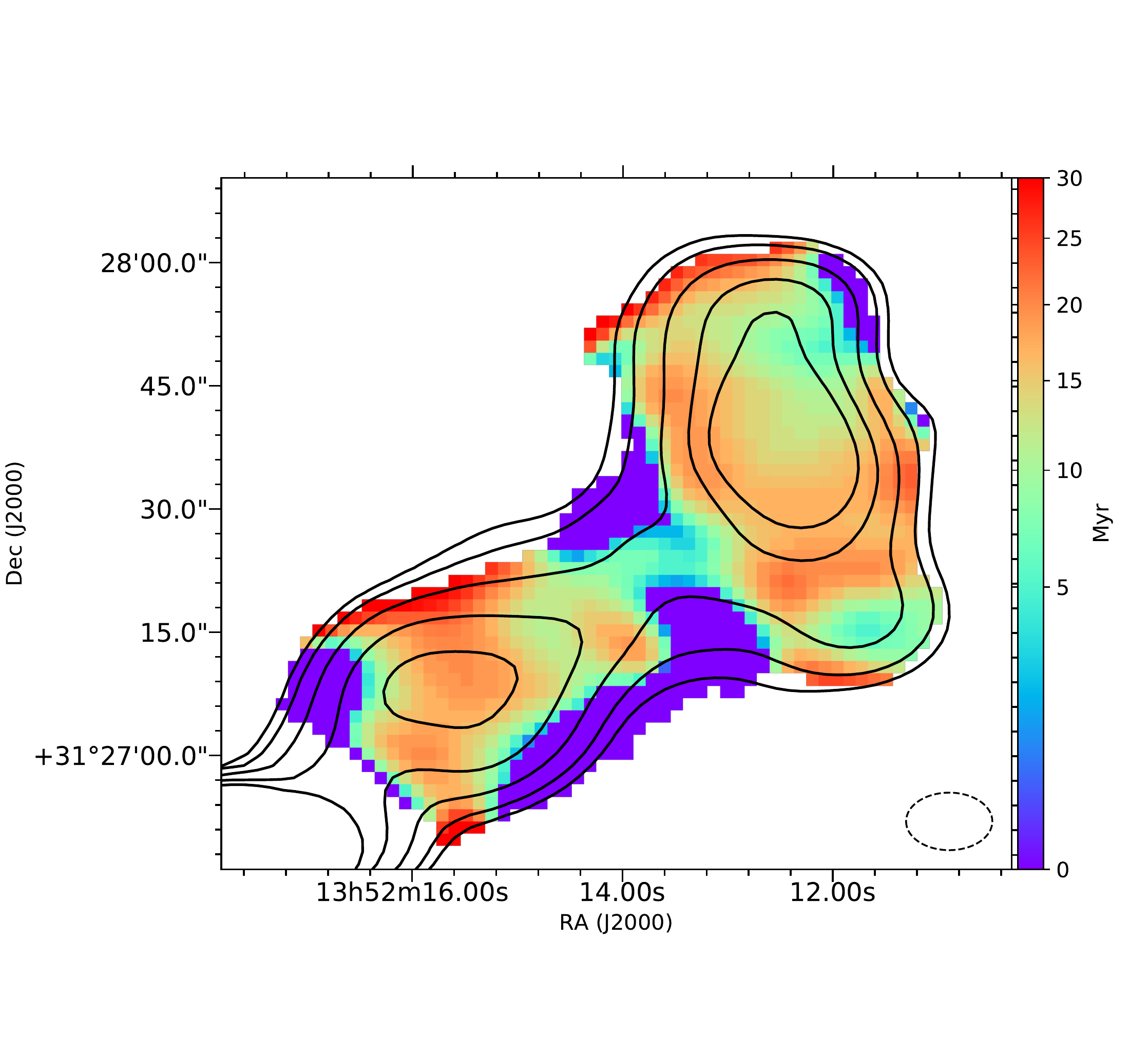}
  \caption{Spectral age map (Myr)}
  \label{fig:jpage}
\end{subfigure}%
\begin{subfigure}{.34\textwidth}
  \centering
  \includegraphics[width=1\linewidth]{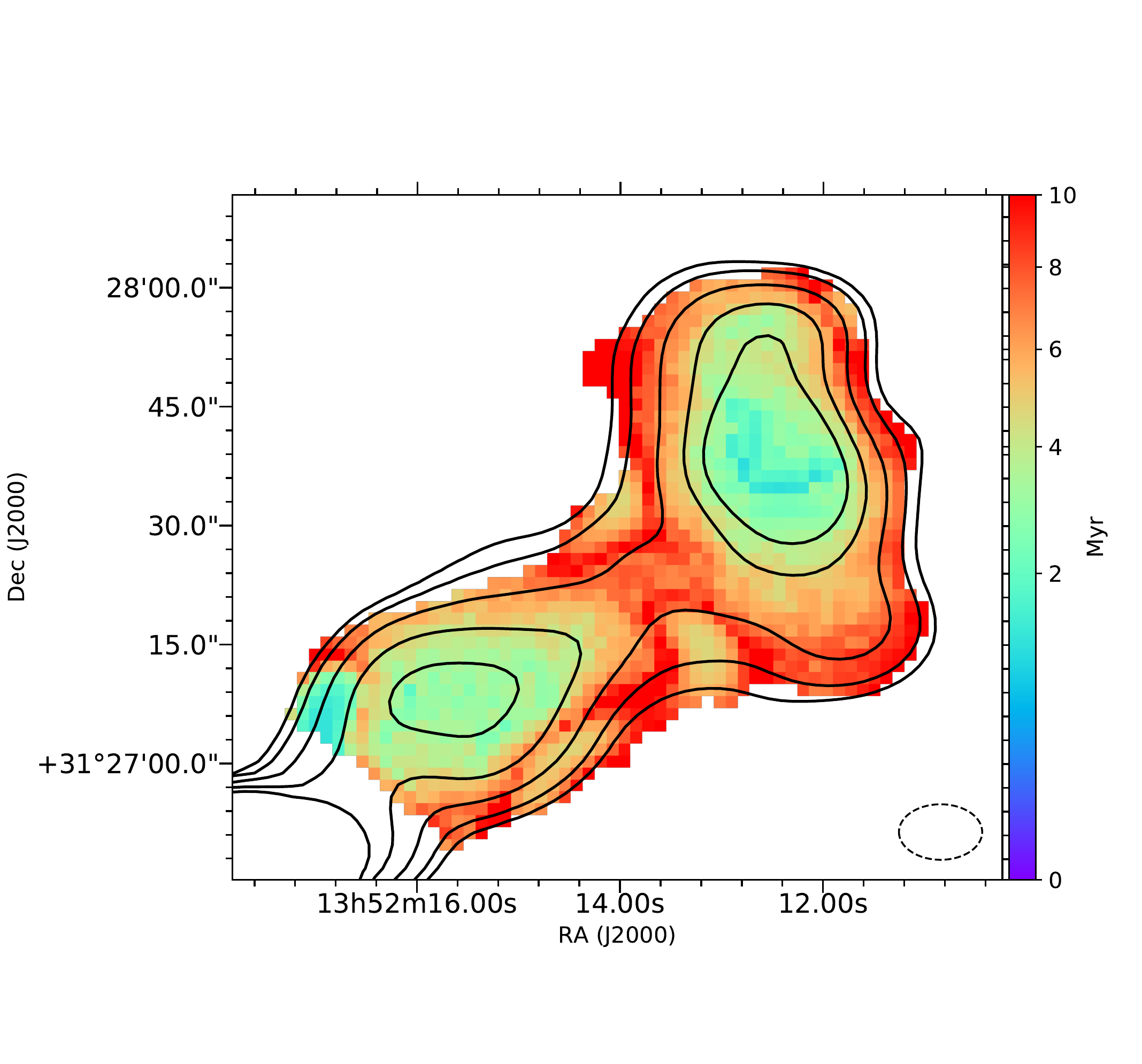}
  \caption{Age error map (Myr)}
  \label{fig:jperror}
\end{subfigure}%
\begin{subfigure}{.34\textwidth}
  \centering
  \includegraphics[width=1\linewidth]{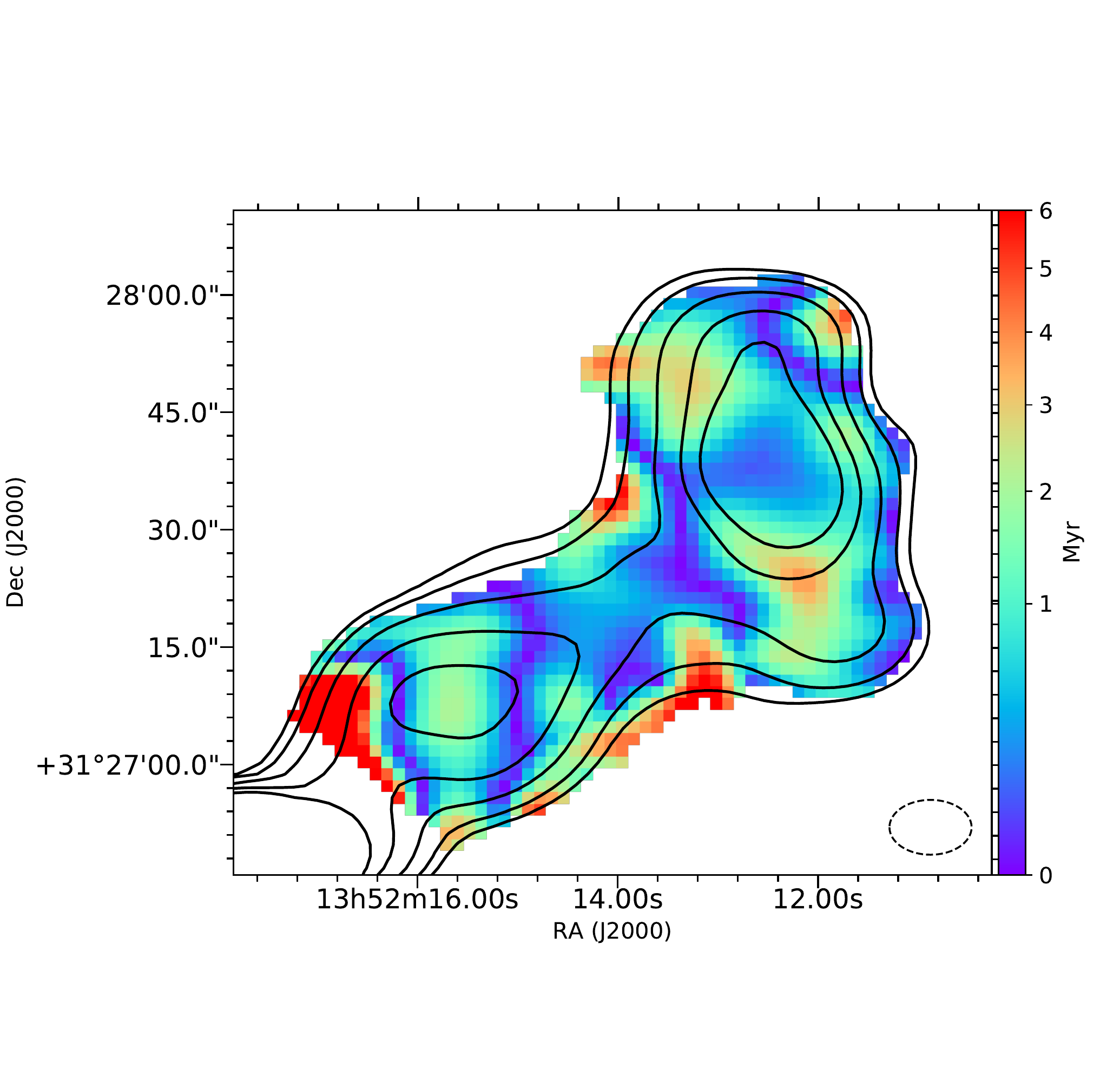}
  \caption{Reduced Chi squared}
  \label{fig:jpchi}
\end{subfigure}
\caption{(a) Spectral age map of the north-western outer lobe of 3C\,293, (b) age error map and (c) reduced $\chi^\mathrm{2}$ map (right) using the JP model. Overlaid are 1360\,MHz emission contours at 10.5\arcsec$\times$7\arcsec and at levels (30, 50, 100, 200, 350)$\times\sigma_\mathrm{RMS}$. }
\label{JPmodelmaps}
\end{figure*}

\begin{table*}
   \centering
  \begin{threeparttable}
       \caption[]{Spectral ageing model fit results for the north-western lobe}
         \label{specage_model}
     \begin{tabular}{ccccccccc}
            \hline
            \hline
            \noalign{\smallskip}
            Model  & Average $\chi^\mathrm{2}_\mathrm{reduced}$ & & & Confidence bins & & & Rejected &Median confidence \\
             & &<68 &68-90 &90-95 & 95-99 & $\geq$ 99 & &\\
            \noalign{\smallskip}
            \hline
            \hline
            \noalign{\smallskip}
    JP & 2.26 & 699 & 399 & 107 & 57 & 44 & No & <68 \\
    Tribble & 1.77 & 783 & 372 & 68 & 47 & 36 & No & <68 \\
            \noalign{\smallskip}
            \hline
            \hline
     \end{tabular}      
    
      \small{`Model' column lists the spectral ageing model fitted to the data. `Average $\chi^\mathrm{2}_\mathrm{reduced}$' column list the average $\chi^\mathrm{2}_\mathrm{reduced}$ of the fit over the entire lobe. `Confidence bins' columns list the number of regions for which the $\chi^\mathrm{2}$ falls within the labelled confidence range. `Rejected' column lists whether the goodness of fit can be rejected over the entire source. `Median confidence' column lists the median confidence level at which the model cannot be rejected.}
     \end{threeparttable} 
   \end{table*}

The shape of the energy spectrum of the electron population can help investigate the evolution of the plasma in radio galaxies. In a system of particles emitting synchrotron radiation under a magnetic field, the energy loss is higher for high energy particles ($dE/dt\propto\nu^\mathrm{2}$) which steepens the spectrum at the higher frequency end. Presence of steep spectra has been used as an indicator for old remnant plasma, devoid of any fresh particle injection. The presence of old plasma along with a region of newer plasma has also been used as evidence for multiple phases of AGN activity. \par
Spectral modelling can be used as a probe for estimating particle ages for an electron population undergoing synchrotron and inverse-Compton losses assuming no new injection. We have used the JP model \citep{Jaffe1973a} here, which assumes that the pitch angle of the magnetic field has a time dependence, which is a realistic assumption for plasma with a lifetime of millions of years. Another model we have used is by \citealt{Tribble1993a}, that allows the electrons to age under a varying magnetic field structure with a Gaussian random distribution (see \citealt{Harwood2013, Hardcastle2013}). A common assumption in these models is that of a single injection event in the past, which can only be physically realistic over small scales of a few kpc.\par

To investigate the spectral ageing of the outer north-western lobe, caused by the natural ageing of plasma due to radiative losses, we have used the BRATS package \citep{Harwood2013,Harwood2015}. We fitted the JP and Tribble models over the entire north-western lobe in a pixel-by-pixel manner to perform a spatially resolved analysis. The assumption of a single injection event and particles being accelerated in the same event can work reasonably well on the scales of our pixel-by-pixel analysis, that is 1.5\arcsec pixel corresponding to 1.3~kpc. We first derived the best fit injection index, $\alpha_\mathrm{inj}$ to be used in the models. For this, we performed a series of fits over the north-western lobe using JP and Tribble models in BRATS, keeping all the other parameters constant and varying $\alpha_\mathrm{inj}$ from 0.5 to 1 with a step size of 0.1. After we found a best fit injection index in this range, we reduced the step size to 0.01 to search around the previous value. The injection index vs reduced $\chi^\mathrm{2}$ for the two models is shown in Figure~\ref{injectionmap}. The plot shows a minimum at $\alpha_\mathrm{inj}$ = 0.61$\pm$0.01 for the two models, with an average reduced $\chi^\mathrm{2}$ of 1.53 for both JP and Tribble models. We use this value as the injection index for our spectral age models.

\begin{table*}[h]
   \centering
  \begin{threeparttable}
       \caption[]{Absorption model fit results for inner lobe regions}
         \label{abs_modelstats}
     \begin{tabular}{cccccccc}
            \hline
            \hline
            \noalign{\smallskip}
            
            Model  & Region  & $a$  & $\alpha$ & $\beta$ & $\nu_\mathrm{p}$ (GHz) & $\chi^\mathrm{2}_\mathrm{reduced}$ \\
             
            \noalign{\smallskip}
            \hline
            \hline
            \noalign{\smallskip}
 
    Homogeneous FFA & E1 & 1.52$\pm$0.05& 0.55$\pm$0.02& ...  & 0.116$\pm$0.009 & 0.36 \\
     \vspace*{10px}
     & W1 & 1.25$\pm$0.02& 0.84$\pm$0.01 & ... & 0.145$\pm$0.004 & 0.11\\

    Internal FFA & E1 & 1.53$\pm$0.05& 0.55$\pm$0.02& ...  & 0.171$\pm$0.015 & 0.35\\
     \vspace*{10px}
    & W1 & 1.25$\pm$0.03& 0.84$\pm$0.01& ...  & 0.225$\pm$0.008 & 0.11\\

    SSA & E1 & 4.14$\pm$0.22 & ... & -2.10$\pm$0.05  & 0.161$\pm$0.010 & 0.38\\
     \vspace*{10px}
    & W1 & 4.96$\pm$0.12 & ... & -2.67$\pm$0.02  & 0.189$\pm$0.003 & 0.09\\
            \noalign{\smallskip}
            \hline
            \hline
     \end{tabular}      
    
      \small{'Model' lists the absorption model fit to the data. 'Region' lists the inner lobe regions, E1 (eastern) or W1 (western). '$a$', '$\alpha$', '$\beta$' and '$\nu_\mathrm{p}$' list the best fit model parameters obtained along with 1\,$\sigma$ errors.} '$\chi^\mathrm{2}_\mathrm{reduced}$' is the reduced chi-square obtained for the fit.
     \end{threeparttable} 
\end{table*}

Using the best fit injection index and the equipartition magnetic field from Sect. 3.3, we fitted the JP and Tribble models over the entire north-western lobe. Spectral age modelling typically requires data at 5 frequencies - 2 below the break to fix the injection index and 3 above the break to measure the curvature due to radiative losses. Although we use only 3 frequencies, our model fit provides reasonable results but likely denote upper limits to the spectral ages. The fitting results are presented in Table~\ref{specage_model} and the spectral age maps obtained with the JP model are shown in Figure~\ref{JPmodelmaps}. The fits obtained using the Tribble model gave results similar to the JP model and therefore we have not shown those maps here. From the table it is clear that none of the models can be rejected at the 68\% confidence level over the entire lobe. The red pixels in the reduced $\chi^\mathrm{2}$ map (Figure~\ref{fig:jpchi}) correspond to the regions that can be rejected at $\geq$ 95\% level. These regions amount to 7\% of the total regions in the lobe. From the spectral age map, we conclude that the spectral ages for most of the north-western lobe lie between 10-20 Myr, especially for the two bright regions in the lobe. The median age for the regions for which the models cannot be rejected with >90\% confidence is 13.1\,Myr.
We find that spectral ages do not show a gradient with distance from the centre. Some regions show no or very little spectral ageing, however, the associated high errors and reduced $\chi^\mathrm{2}$ values reduce our confidence in the model fits over these regions. \par
For the central regions, we use the following equation \citep{Kardashev1962a,Murgia2011DyingClusters} - 
\begin{ceqn}
\begin{align}
t_\mathrm{s} = 1590 \frac{B^{0.5}_\mathrm{eq}}{(B^{2}_\mathrm{eq}+B^{2}_\mathrm{CMB})\sqrt{\nu_\mathrm{b}(1+z)}},
\end{align}
\end{ceqn}

where $t_\mathrm{s}$ is in Myr, the magnetic fields $B_\mathrm{eq}$ and B$_\mathrm{CMB}$=3.25$\times$(1+z)$^{2}$ are in $\mathrm{\mu}$G, and the break frequency $\nu_\mathrm{b}$ is in GHz. Assuming the break frequency to be the highest frequency of our observations, that is 8.45\,GHz, we estimate upper limits on the spectral ages -
$\lesssim$0.15$\pm$0.01\,Myr for E1, $\lesssim$0.17$\pm$0.02\,Myr for W1, $\lesssim$0.27$\pm$0.04\,Myr for E2 and $\lesssim$0.23$\pm$0.03\,Myr for W2. \citet{Machalski2016a} estimate a dynamical age of $\sim$0.3\,Myr for the central region. Their estimate is similar to ours within errors, but they do not include the absorption in the spectra, which makes it difficult to compare the two ages. The difference from their value could also be due to the different in the magnetic field strength estimated between the two studies.
\subsubsection{Absorption models for inner lobes}

The spectra for the inner lobe regions E1 and W1, are relatively flat in the 144-1360\,MHz range and below the theoretical limit for injection index ($\alpha_\mathrm{inj}$ = 0.5). This suggests that the spectra peaks in between 144 and 1360\,MHz which implies the presence of low frequency absorption. 
An absorbed spectrum is seen compact steep spectrum (CSS) or gigahertz peaked spectrum (GPS) radio galaxies. This peak (or turnover) is attributed to the absorption of synchrotron radiation in the source, which is broadly a manifestation of either synchrotron self absorption (SSA) or free-free absorption (FFA) \citep{Kellermann1966,Tingay2003An1718-649,Callingham2015a}.\par

The best way to identify the absorption mechanism is to measure the slope of the spectrum below the peak frequency: since FFA shows a much steeper index ($\alpha\lesssim-2.5$) than SSA. Given the lack of more high resolution observations below 144 MHz, we cannot directly estimate the spectral index below the peak. Therefore, we have fitted various absorption models to the spectra for E1 and W1, similar to \cite{Callingham2015a} and used the derived fit parameters to discriminate between these models in Section 4.1.1.
\begin{enumerate}
    \item Synchrotron Self Absorption (SSA) -  This is a standard SSA model that assumes self absorption from a synchrotron emitting plasma due to the scattering of the emitted synchrotron photons by the relativistic electrons. The absorption cross section is higher for longer wavelengths and therefore as the observing frequency increases, photons emerge from the deeper regions of the source until the optically thin regime is reached. This model for a synchrotron emitting homogeneous plasma is given by - 
    \begin{ceqn}
    \begin{align}
    S_\mathrm{\nu} = a \left(\frac{\nu}{\nu_\mathrm{p}}\right)^\mathrm{\frac{\beta+1}{2}} \left(\frac{1-e^\mathrm{-\tau}}{\tau}\right),
    \end{align}
    where 
    \begin{align}
     \tau = \left(\frac{\nu}{\nu_\mathrm{p}}\right)^\mathrm{\frac{\beta-4}{2}}.
    \end{align}
    \end{ceqn}
    We note that $\nu_\mathrm{p}$ is the frequency at which the source becomes optically thick ($\tau$=1), $\beta$ is the power law index of the electron energy distribution related to the synchrotron spectral index as $\alpha= -(\beta+1)/2$ and $\tau$ is the optical depth.
    
    \item Homogeneous Free Free Absorption (FFA) - This model assumes that the attenuation of radiation is caused by an external homogeneous ionised screen around the relativistic plasma emitting the synchrotron spectrum. The free-free absorbed spectrum is then given as - 
    \begin{ceqn}
    \begin{align}
    S_\mathrm{\nu} = a \nu^\mathrm{-\alpha} e^\mathrm{-\tau_\mathrm{\nu}},
    \end{align}
    \end{ceqn}

    where $a$ and $\alpha$ are the amplitude and spectral index of the intrinsic synchrotron spectrum, and $\tau_\mathrm{\nu}$ is the optical depth. The optical depth is parameterised as $\tau_\mathrm{\nu}=(\nu/\nu_\mathrm{p})^\mathrm{-2.1}$, where $\nu_\mathrm{p}$ is the frequency at which the optical depth is unity.
    \item Internal FFA - In this case, the absorbing ionised medium is mixed with relativistic electrons that produce the synchrotron spectrum. This model is given as - 
    \begin{ceqn}
    \begin{align}
    S_\mathrm{\nu} = a \nu^\mathrm{-\alpha} \left(\frac{1-e^\mathrm{-\tau}}{\tau}\right).
    \end{align}
    \end{ceqn}
\end{enumerate}

To fit these models to the inner lobes, we used the integrated flux densities and their errors, extracted from regions E1 and W1 and summarised in Table~\ref{region_summary}. The absorption models were then fitted to the data using the SciPy python package which utilises the Levenberg–Marquardt optimisation method. The resulting model fits are shown in Figure~\ref{abs_models} and the fit parameters are summarised in Table~\ref{abs_modelstats}. Both SSA and FFA models provide similar quality fits to our data. We find that for all three models, the $\nu_\mathrm{p}$ frequency, the frequency at which optical depth is 1, is systematically higher for the western lobe than the eastern lobe, with a 3$\sigma$ significance. To test the robustness of our analysis, we performed the same fitting procedure to integrated fluxes from regions of different sizes in the inner lobes and found that our results were in agreement. The results of the models and their implications for the inner lobes will be discussed further in Sect. 4.1.

\section{Discussion}
The evolution of the radio emission in 3C\,293 is complex and our results confirm this. Here we discuss the results of our spectral analysis going from the inner to the outer scale emission in the context of understanding the life-cycle and evolutionary scenarios for 3C\,293.

\subsection{Interplay of radio plasma and gas in the central region}
3C\,293 shows bright radio inner lobes surrounded by diffuse emission, with a total linear extent of $\sim$4.5\,kpc. These compact components have flat low frequency spectra (Section 3.5) and are surrounded by a dense and rich ISM. Previous studies have found evidence for jet-ISM interaction in the galaxy. In their study of ionised gas kinematics, \citet{Mahony2016} found jet-driven outflows and disturbed gas throughout the central region of the host galaxy. Outflows of warm gas have also been observed in the galaxy by \citet{Emonts2005}. In the neutral gas, \citet{Morganti2003} found fast outflows, up to 1400 \kms{}, in central regions and also suggest that it is driven by the interaction between the radio jets and the ISM. More recently, \citet{Schulz2021} have also used global VLBI to map the HI outflow in the galaxy. \citet{Lanz2015} studied the galaxy in X-ray and found more evidence for jet-ISM interaction. They concluded that the X-ray emission from the central region is caused by shock heating of the gas by this interaction. \citet{Massaro2010} also observed X-ray emission from the radio jets. Here, we explore what the spectral properties of these components tell us about the system.

\subsubsection{What causes absorption in the inner lobes?}

\begin{figure}
   \includegraphics[width=\columnwidth,height=1.15\columnwidth]{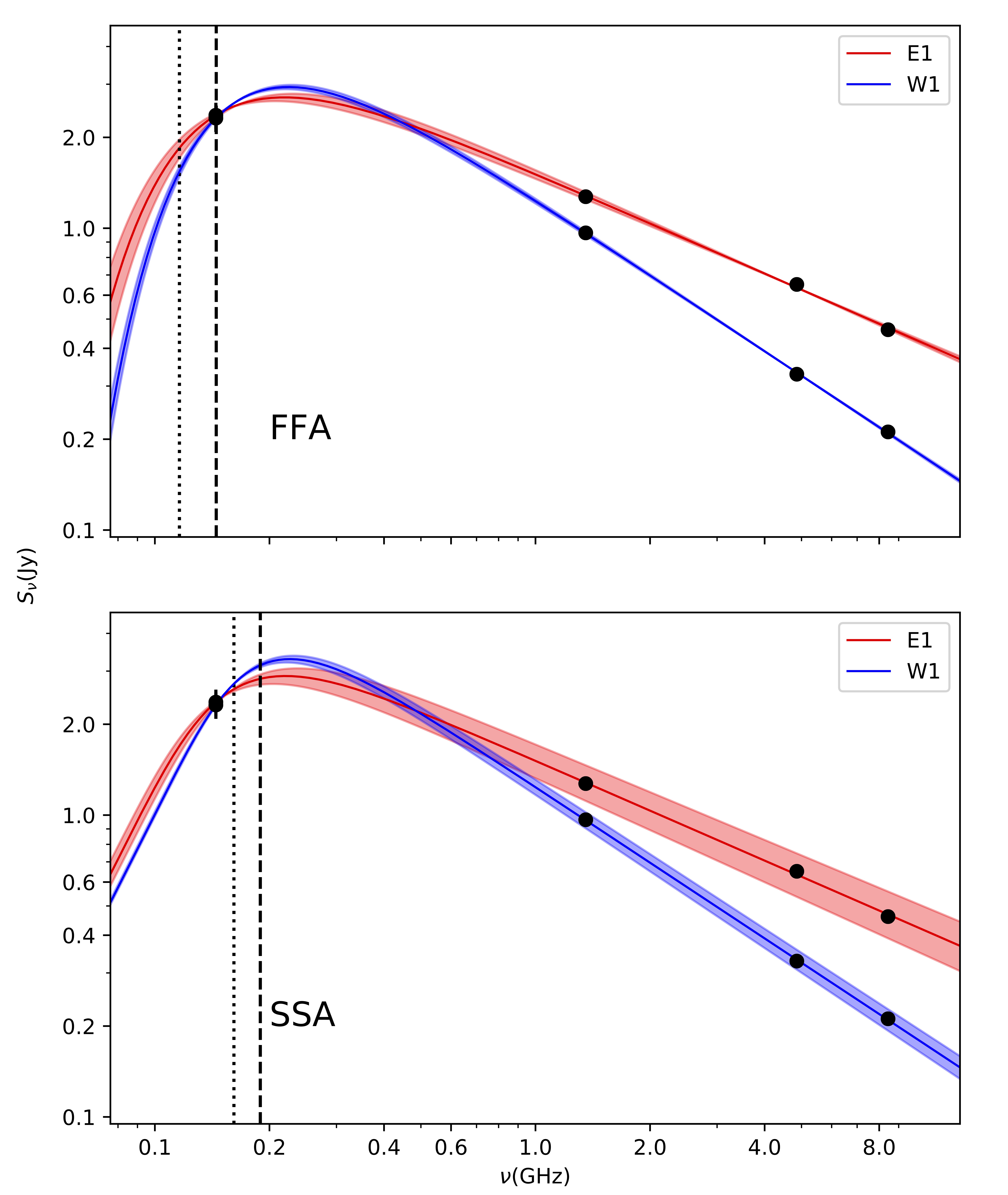}
   \caption{Best fits of Homogeneous FFA (top panel) and SSA (bottom panel) absorption models to the E1 (eastern) and W1 (western) inner lobes regions of 3C\,293. A peak in the spectrum can be seen at 220-230\,MHz. All models provide similar quality fits to the data. The vertical lines show the $\nu_{p}$ frequency, for E1 (dotted) and W1 (dashed). $\nu_{p}$ is the frequency at which optical depth $\tau$=1, an is therefore a measure of the optical depth. From the model fits, $\nu_{p}$ for W1 is systematically higher than for E1.}
    \label{abs_models}
\end{figure}
As discussed in Section 3.4.2, we have used our model fit parameters to discriminate between the SSA and FFA models. \par
\vspace{2.5mm}
(i) SSA :
From the fit parameters obtained for SSA in Sect 3.4.2 (see Table~\ref{abs_modelstats}) we derive a peak frequency of 220\,MHz and 226\,MHz for E1 and W1, respectively. If SSA is the origin of the turnover, we can now relate these parameters to the properties of the source. For pure SSA, in a homogeneous synchrotron self-absorbed radio source, the magnetic field $B$ is related to the peak frequency $\nu_\mathrm{peak}$ and the flux density at the peak frequency $S_\mathrm{max}$. Under an equipartition assumption between magnetic field and the electrons, the peak frequency is given by \citep{Kellermann1981,Callingham2015a}
\begin{ceqn}

\begin{align}
    \nu_\mathrm{peak} \sim 8.1 B^\mathrm{1/5} S_\mathrm{max}^\mathrm{2/5} \theta^\mathrm{-4/5} (1+z)^\mathrm{1/5},
\end{align}
\end{ceqn}
\noindent
where $\nu_\mathrm{peak}$ is in\,GHz, $B$ is the magnetic field in G, $S_\mathrm{max}$ is the peak flux density at the peak frequency in\,Jy and $\theta$ is the angular size of the source in mas. The relation depends very strongly on observables such as angular size and peak frequency, which are very difficult to estimate accurately. Therefore, the computed magnetic field strengths should be treated with care.\par
We use $\theta$ = 692.8 mas, which is estimated using $\theta=\sqrt{\theta_\mathrm{1}\theta_\mathrm{2}}$, where $\theta_\mathrm{1}$ = 800 mas and $\theta_\mathrm{2}$ = 600 mas are the two dimensions of the regions. We estimate a magnetic field of 4.9$\times10^\mathrm{2}$ G for E1 and 4.3$\times10^\mathrm{2}$ G for W1 region. Using more accurate sizes of the inner lobes from the 30 mas high resolution images from \citet{Beswick2004}, equal to $\sim$400 mas, we estimate magnetic fields 2-8 times lower. However, these values are significantly higher than our equipartition estimates in Section 3.3, by $\sim$10$^{7}$. Our estimates from SSA are also at least $10^\mathrm{3}$ times higher than the typical values found in GPS sources, which are in the range 5-100 mG \citep{ODea1998b,Orienti2008}. For CSS sources of similar sizes and peak frequencies from \citet{ODea1998b}, such as 3C\,48, 3C\,138 and 3C\,147, we estimate magnetic field strengths of $\approx$ 4-15 mG using equation 6.\par
For a sample of young radio sources, \citet{Orienti2008} find good agreement between equipartition magnetic fields and those derived using the peak frequency assuming SSA thus concluding that the turnover in their spectra is probably due to SSA. However, for a couple of their sources (J0428+3259 and J1511+0518), the magnetic fields from SSA relation were significantly higher than the equipartition values and they suggest that for these sources, a more likely explanation is that the spectral peak is caused by free-free absorption. \par
The unrealistically high magnetic field strengths estimated assuming SSA cannot sustain synchrotron emission in the system (loss lifetimes of $< 3\times$ $10^{-4}$ yr), even for the light travel time across the region. This tells us that it is unlikely that SSA is the dominant or only absorption mechanism in the inner lobes of 3C\,293. \par
\vspace{2.5mm}
(ii) FFA : Free-free absorption requires the presence of a dense optical line-emission medium and strong depolarisation of the radio source. Multiple studies have found a dense medium in the central few kpc of 3C\,293 \citep{Emonts2005,Labiano2014,Mahony2016} and \citet{Akujor1996b} have confirmed strong depolarisation in the source which may be caused by the gaseous disc revealed by optical emission lines. \par
From our best fit results in Table~\ref{abs_modelstats} (also see vertical lines in Figure~\ref{abs_models}), we find that the optical depth is systematically higher for W1 than E1 with a 3\,$\sigma$ significance. 
Over the years, resolved absorption studies of CSS/GPS sources have found that a difference between the optical depths of the lobes is due to the larger path length in the ionised medium along the line of sight for the emission from the receding lobe, for example OQ\,208 \citep{Kameno2000}, NGC\,1052 \citep{Kameno2001}, 3C\,84 \citep{Vermeulen1994,Walker1994,Walker1999}, NGC\,4261 \citep{Jones2000,Jones2001} and NGC\,6251 \citep{Sudou2001} (and also see Figure 6 of \citealt{Kameno2000}). The asymmetry in the derived optical depths is consistent with the current understanding of the orientation of the inner jets \citep{Beswick2004,Mahony2016}.\par

In their study of FFA in GPS sources, \citet{Kameno2005} found that the ratio of the optical depths of the lobes ($\tau_\mathrm{A}/\tau_\mathrm{B}$) is $<$ 5 for sources where the line of sight is nearly perpendicular to the jet axis. We find an optical depth ratio of $\sim$1.6 between the inner lobes, which shows that this is not a case of highly asymmetric FFA.\par

\citet{Beswick2004} in their high resolution study of the inner lobes also attribute the steeper index of the western jet to the presence of FFA. In the case of FFA from an external homogeneous ionised screen, the emission measure of the absorbing medium is given as \citep{ODea1998b}- 
\begin{ceqn}
\begin{align}
    n_\mathrm{e}^\mathrm{2} L = 3.05 \times 10^\mathrm{6} \tau \left( \frac{T}{10^\mathrm{4}K}\right)^\mathrm{1.35} \left( \frac{\nu}{1\,GHz}\right)^\mathrm{2.1} \mathrm{cm}^\mathrm{-6} \mathrm{pc},
\end{align}
\end{ceqn}
\noindent
where $n_\mathrm{e}$ is the electron density in cm$^\mathrm{-3}$, $L$ is the path length in pc, $\tau$ is the optical depth at frequency $\nu$ in\,GHz. FFA in galaxies is generally attributed to the Narrow Line Region (NLR) clouds around the radio jets and a filling factor $f$ has been included in the path length to account for the clumpy nature of the absorbing medium. We estimate a value of $f=4.3\times 10^\mathrm{-6}$ using a narrow line H$\beta$ luminosity of L(H$\beta)_\mathrm{narrow}$ = 1.3$\times$10$^\mathrm{40}$ ergs s$^\mathrm{-1}$ and $n_\mathrm{e}$ = 200 cm$^\mathrm{-3}$ from \citet{Emonts2005}. Using these parameters, we estimate a path length of $\approx$50 pc for E1 and $\approx$80 pc for W1. These path lengths are easily achievable given the evidence for narrow line region clouds out to a few kpc and a dense ISM \citep{Emonts2005}.\par
In their study of alternative FFA models, \citet{Bicknell1997} found that the optical depth in a free-free absorption screen will vary with radius if the medium around the radio source is affected by shocks. Previous studies have confirmed the presence of shocks in the medium surrounding inner lobes of 3C\,293 \citep{Lanz2015,Mahony2016}. Thus it is possible that the difference we see in the optical depths for the two lobes is a consequence of inhomogeneous FFA. However, given the limitations of our current data sets, we cannot investigate this effect further and would need more high resolution observations below the peak frequency to discriminate between the two FFA models. \par
In case of FFA being responsible for the spectral turnover, $\nu_\mathrm{peak}$ due to SSA would be lower with a higher peak flux density $S_\mathrm{max}$ than observed. The magnetic field from equation 5 depends on these observables as $B\propto \nu_\mathrm{peak}^\mathrm{5}S_\mathrm{max}^\mathrm{-2}$, and therefore the true field strength would be less than estimated using the observed peak and attributing it to SSA. This would explain the unrealistically high magnetic fields estimated from equation 5. We conclude that the most realistic situation is that FFA is dominant but is likely not the sole absorption mechanism and SSA also contributes a component to the absorbed spectrum.
\par

\subsubsection{Are the inner lobes a young source?}
\citet{Akujor1996b} have speculated the presence of a CSS source in the inner lobes which are $\sim$2\,kpc in size and contribute a significant fraction to the total flux density. Even higher (mas) resolution images of the inner lobes at 1.4\,GHz and 4.5\,GHz have found bright jet emission and radio knots in this region \citep{Akujor1996b,Beswick2004} towards the end of the jet, which tells us that the inner lobes are jet dominated. \par

Multiple studies of CSS and GPS sources have found a correlation between the linear size of the source and the peak frequency of the spectrum, in both high power \citep{Bicknell1997,ODea1997} and low power \citep{DeVries2009} sources. Although this correlation has been explained in terms of SSA \citep{Snellen2000,Fanti2009}, FFA models with absorption due to an inhomogeneous medium have also been able to recreate the relation \citep{Bicknell1997,Kuncic1997InducedSources}. However, FFA via a homogeneous external medium cannot replicate such a relationship \citep{ODea1998b}. This correlation is given by -
\begin{ceqn}
\begin{align}
    \log(\nu) = -0.21(\pm0.05) - 0.65(\pm0.05) \log(\mathrm{L}_{PLS}),
\end{align}
\end{ceqn}
\noindent
where $\mathrm{L}_{PLS}$ is the linear size in kpc and $\nu$ is the peak frequency in GHz. We deproject our linear size by using the viewing angle with respect to the jet axis, estimated to be 55\degree and 75\degree by \citet{Beswick2004} and \citet{Machalski2016a}, respectively. This gives a deprojected physical size of 2.1-2.4\,kpc. The minimum peak frequency corresponding to this is $349^\mathrm{+26}_\mathrm{-24}$MHz. This is significantly higher than the rest frame peak frequencies of 230-240\,MHz for the inner lobes that we estimate from the absorption models in Section 3.4.2. This suggests that the inner lobes are strongly interacting with and prevented from expanding by, the rich surrounding medium of the host galaxy, also found by other studies mentioned before.

\par
As discussed in Sect. 3.4.1, we obtained ages of $\lesssim$0.15$\pm$0.01\,Myr for E1 and $\lesssim$0.17$\pm$0.02\,Myr for W1. However, for 3C\,293, \citet{Emonts2005} have estimated that jet induced outflows, for a constant velocity, must have been driven for $\sim$1 Myr in order to obtain the total outflow mass of HI and ionised gas. Our spectral ages are much lower than their estimate, but it is possible that older phases of jet activity have contributed to the outflow mass or that the velocity of the gas has decreased with time and expansion of the outflow. The spectral ages are in agreement with jet dominated CSS sources of similar sizes found by \citet{Murgia2002a}, who found ages in the range of 10$^\mathrm{3}$ to 10$^\mathrm{4}$ yr, and attributed them to the bright jet components. This tells us that the spectral ages we estimate do not represent the actual source age, but the permanence time of the electrons in the bright compact jet components which dominate the flux density and where the electron reacceleration occurs. This again supports a scenario where jet-ISM interaction impedes the flow of plasma. \par

It is unlikely that these lobes are just strongly interacting and not young, that is they are as old as the large-scale radio galaxy. This scenario also supports FFA by the rich surrounding medium being the dominant absorption mechanism, as we discussed in Sect. 4.1.1. Other properties such as depolarisation at high frequencies, Faraday dispersion of $\Delta<1200$ cm$^\mathrm{-3}\mu$G pc and minimum pressures in the inner lobes of 3C\,293 are also typical of CSSs \citep{Akujor1996b}. \par

\subsubsection{What is the origin of diffuse emission around inner lobes?}
The morphology of the regions adjacent to the inner lobes (W2 and E2) is intriguing. There is an abrupt increase in the width of the emission and reduction in the surface brightness compared to the inner lobes. The $\alpha^\mathrm{144}_\mathrm{1360}$ spectral index of the diffuse emission also shows a steepening with respect to the inner lobes. Assuming that the emission from E2 and W2 is optically thin, the spectral age is
$\lesssim$0.27$\pm$0.04\,Myr for E2 and $\lesssim$0.23$\pm$0.03\,Myr for W2, as discussed in Sect. 3.4.1.
 
The steeper spectra and the morphology of the diffuse emission regions suggest that they could be from an older phase of activity. These diffuse regions representing an older phase of activity was also suggested by the spectral study of \citet{Akujor1996b}.\par 
Another plausible scenario is that the diffuse regions are formed by the leaked plasma from the radio jet as it propagates through the galaxy's disk. In their study of simulated interaction between a jet and the galaxy's disk, \citet{Mukherjee2018b} found that a jet inclined to the plane of the disc will interact strongly with the gas in the disc and will not immediately clear the medium and move out. It will deflect and decelerate, with plasma leaking out along the path of least resistance (see their Figure 14). It is likely that this is the case for 3C\,293, where the jets are inclined with respect to the host galaxy's disc \citep{Floyd2005,Labiano2014}. Leakage from the decelerated plasma that moves out of the disk, in directions perpendicular to the jet flow, would form the lower surface brightness diffuse emission we see in these regions. Another prediction of the simulation from \citet{Mukherjee2018b} is the presence of outflows, as seen in 3C\,293 and described before.

\subsection{Evolution of the outer lobes}
From Figure~\ref{JPmodelmaps}, we observe spectral ages varying typically from $\approx$10-20 Myr over the two bright regions (also marked by O1 and O2) of the north-western outer lobe, with a median age of 13.1\,Myr. \citet{Machalski2016a} have estimated the dynamical age to be $\sim$62 Myr, which gives a dynamical to spectral age ratio between $\sim$3 and $\sim$6 for most of the lobe.

It is possible that this difference is a result of the equipartition assumption for the magnetic field. Indeed, in their study of spectral and dynamical ages, \citet{Mahatma2020} found that the equipartition assumption can underestimate the spectral age by factors of up to $\sim$20 and that the actual magnetic field strength in FRII galaxies is sub-equipartition. \par

In order to understand the evolution of the outer lobes, there are two important spectral properties to note (Section 3.2.1). The first property is the small curvature from 144\,MHz to 4850\,MHz. We do not see any sign of an ultra-steep spectral index ($\alpha^\mathrm{144}_\mathrm{1360}$ or $\alpha^\mathrm{1360}_\mathrm{4850}$ > 1.2) up to 5\,GHz.
The second property is lack of a dependence of the spectral index on distance from the centre, which seems to be uniformly distributed on a scale of about 10\,kpc throughout the 90\,kpc lobe.\par

The absence of any ultra steep spectral index suggests that the outer lobes are not remnants. The lack of a spectral gradient is contrary to the trend expected, as plasma in different regions would have different ages and would be expected to show different curvatures in their spectra.
This suggests that the plasma in the lobe is highly turbulent which reaccelerates old electrons (due to shock acceleration) and mixes different electron populations. The diffuse morphology of the north-western lobe also suggests that the flow is not very well collimated and more turbulent.\par

\par
Such scenarios have been suggested to be active in galaxies with similar properties (bright resolved inner kpc region and outer, low-surface brightness, diffuse lobes), for example Centaurus A \citep{McKinley2018}, B2~0258+35 \citep{Brienza2018} and more recently NGC 3998 \citep{Sridhar2020}. For Centaurus A, \citet{Eilek2014a} have estimated that in case of a lack of injection of fresh plasma, the turbulence would last a few tens of Myr ($\sim$30 Myr) after which the lobes will fade away. Keeping in mind that the size of 3C\,293 ($\sim$220\,kpc) is much smaller than that of Centaurus A ($\sim$500\,kpc), it is likely that these processes could be active in the outer lobes of 3C\,293. The presence of shocks in the outer lobes of 3C\,293 has been suggested by \citet{Lanz2015}, who in their study of X-ray emission from 3C\,293, concluded that the presence of shell-like morphological features in the outer lobes and their possible thermal origin mean that they could be associated with bowshocks, that heat the gas to X-ray temperatures.

Keeping the spectral properties in mind, we propose two scenarios for the evolution of the outer lobes of 3C\,293. One scenario here does not preclude the other.  
\begin{enumerate}

    \item The outer lobes represent older phases of activity. In this scenario, the interruption would have happened for a very short time, possibly only a few Myr ago ($\sim$0.7\,Myr from \citealt{Machalski2016a}), which would explain the lack of an ultra steep spectral index anywhere in the lobe. As mentioned in Section 3.1.1, the bright emission at the end of the lobe (O2) has peculiar morphology which could mean that it is from an older phase of activity, or represents a variable and intense phase of activity. The lack of a spectral gradient would be due to the shocks caused by the newer jet material expanding (O1) into the older lobe material (O2), supersonically. This would also explain why such spectral properties are not observed in other restarted radio galaxies which are hydrodynamically not as complex as a galaxy with several episodes of jet restarting and are powered by a single constant flow. In this scenario, the young CSS source in the inner lobes of 3C\,293 would be formed in a newer phase of activity.\par
  
    \vspace{3mm}
    
    \item The AGN has not switched off completely, and the outer lobes are still fuelled by the centre. The jet flow is intermittent due to a strong interaction between the jets and the dense ISM. In this scenario, the fuelling of the lobes by the centre would explain the lack of a steep spectra or a strong curvature. The strong interaction of the jets with the dense ISM would disrupt the flow and trap the plasma, building up a deposition of energy until the plasma finally breaks through and expands in the outer lobes. The decollimation of the jet flow due to the interaction would result in a turbulent flow of plasma in the outer lobe. This would also explain the abrupt change in the surface brightness of the jets outside the nuclear region. However, in this scenario, the fuelling of the outer lobes by the centre would be needed to maintain the turbulence that keeps the spectra from steepening. We do not see any direct extension of the diffuse emission in the centre to the outer lobes, therefore we cannot confirm the presence of an open plasma transport channel to these lobes. It is possible that we do not have enough sensitivity in our sub-arcsecond images to recover such lower surface brightness emission and therefore, we cannot rule out the presence of such a plasma transport channel.  
\end{enumerate}    
As mentioned before in Section 3.2.1, the spectral index distribution in the south-eastern lobe suggests multiple episodes of jet activity, similar to the north-western lobe. The presence of similar X-ray features in this lobe \citep{Lanz2015} also suggests that similar processes are active here.\par

\subsection{3C\,293 and other similar galaxies}
3C\,293 appears to belong to the growing group of radio galaxies, that were classified as restarted based on their morphology (bright central region and low surface brightness diffuse lobes) and have been found to show no curvature in the radio spectra of their lobes due to ageing. Other examples of such galaxies with a prominent central (kpc) emission and extended (much lower surface brightness) emission include Centaurus~A, NGC\,3998 and B2~0258+35. \par
In Centaurus A, \citet{McKinley2018} found a uniform index of $\alpha^\mathrm{154}_\mathrm{2300} \sim 0.8$ spread across the lobes (see their Figure 4). They argued that particle re-acceleration due to turbulence powered by the jets of the central engine is responsible for the uniform distribution of the spectral index. Presence of large-scale channels connecting the outer lobes to the centre was confirmed by \citet{Morganti1999,McKinley2018} which supports the scenario where the turbulence in the outer lobes is being maintained by the fuelling from the centre. \citet{Brienza2018} found
similar properties of the spectral index in B2\,0258+35. They proposed that scenarios such as jet flow disruption or episodic activity with a short interruption could power in situ particle reacceleration and/or adiabatic compression that would prevent spectral steepening. More recently \citet{Sridhar2020}, in their study of NGC\,3998, also found a spectral index of $\alpha^\mathrm{147}_\mathrm{1400}=0.6$ uniformly spread out over the lobes (see their Figure 4). They propose similar scenarios of sputtering activity and jet flow disruption. \par

\citet{Jurlin2020} found a comparable fraction of restarted and remnant radio galaxies in their sample, and suggested that activity can restart in galaxies after a short remnant phase. The discovery of the above mentioned group of galaxies is in agreement with their result. \citet{Shabala2020a} also found that a model with power-law distribution of the ages of radio galaxies was able to reproduce the observed fraction of restarted and
remnant radio galaxies much better than a model with a constant age for all sources. They found that the best fit model was obtained with a higher fraction of short lived sources (<100\,Myr). The discovery of the above mentioned group of galaxies and the present study of 3C\,293 are in line with these conclusions. This sub-group of galaxies has also been found to be gas rich, which sometimes in combination with the presence of a merger, may play a role in providing the right condition for a fast duty cycle. 
\par
In the future, more occurrences of radio sources in similar conditions could be identified using spatially resolved spectral studies down to low frequencies combined with the information about the gas medium of the host galaxy.
LOFAR would provide the low frequency information, where injection and absorption effects are relevant and help estimate the age and evolutionary stage of the galaxy. APERture Tile In Focus (Apertif) phased-array feed (PAF) system \citep{Adams2019} would provide the HI kinematics of the host galaxy along with the higher frequency radio continuum. WHT Enhanced Area Velocity Explorer (WEAVE; \citealt{Dalton2016}) would be needed to probe the ionised gas properties and estimate accurate redshifts for these galaxies. Exploiting the synergy between these instruments would allow us to search for a connection between the evolutionary stage and the gas properties of these galaxies and give new insights into their life-cycle.\par

\section{Summary and future prospects}
This is the first time that the spectral properties of 3C\,293 and their spatial distribution has been explored on both large and small scales down to frequencies where the breaks become apparent. 3C\,293 has long been classified as a restarted galaxy, although our detailed analysis has revealed that it is not a typical restarted galaxy with a new phase of activity embedded in diffuse emission lobes with properties of remnant plasma. We find the following:
\begin{enumerate}
    \item We have observed for the first time, absorption in the inner lobes of 3C\,293 with a peak frequency of $\sim$230-240\,MHz. Free-free absorption from the NLR in the rich surrounding medium of the host galaxy is likely the dominant absorption mechanism.
    \item From the age, size and presence of a turnover in the spectrum of the inner lobes of 3C\,293, we conclude that they are a young CSS source whose growth is affected by the dense surrounding medium. This confirms that 3C\,293 is indeed a restarted galaxy. The spectral ages of $\lesssim$0.15\,Myr \& $\lesssim$0.17\,Myr represent the permanence time of the electrons in the bright compact jet components.
    \item The diffuse emission seen in the centre of 3C\,293 is likely formed by the leakage of radio plasma from the jet that is deflected and decelerated by interaction with the galaxy's medium. 
    \item The spectral properties of the outer lobes are a result of either one or maybe both of the two scenarios - multiple episodes of jet activity and turbulent jet flow due to disruption by strong jet-ISM interaction. Any interruption of jet activity has happened only a few Myr ago and the lobes are not made of remnant plasma, they are still alive. Shock powered turbulence has kept the spectra from steepening and the spectral distribution uniform. Overall, also considering the young CSS source in the centre, we conclude that 3C293 has had at least two to three epochs of activity.  
\end{enumerate}

Although 3C\,293 has outer lobes that show spectral properties of active fuelling (and are hence alive), the presence of a young CSS source in the centre has confirmed that it is indeed a restarted galaxy. Finding similar sources and correctly categorising them separately from typical restarted galaxies is important, since shorter interruption time periods would affect the overall understanding of AGN life cycle timescales, which is a crucial input for AGN feedback models.\par
The ability to spatially resolve the spectral properties of the small and large-scale emission down to low frequencies can be of great importance, as shown by our study, in understanding the evolution of the radio AGN over its various life cycles and can reveal multiple epochs of activity, that single frequency morphology studies cannot do. Although resolved studies of absorption have been carried out before, they were only possible for sources with spectral peaks at GHz frequencies, due to the previously limited resolving power of instruments before the ILT. With the International stations of ILT providing sub-arcsecond resolution at 144\,MHz (HBA), and in the future 57\,MHz (LBA), we can now perform detailed studies of low frequency peaked sources and understand the different mechanisms that play out. Studying the ILT statistics of radio luminosity along the jets of the huge number of radio galaxies and quasars could also provide a unique diagnostic of the timeline of nuclear activity in long-living radio-loud active galaxies and quasars. This study has shown that it is crucial to quantify and probe properties at both large and small scales in order to gain a complete understanding of the different types of restarted galaxies. This opens up a new and exciting low frequency window into understanding galaxies that have eluded us at higher frequencies. 
    
\begin{acknowledgements}
LKM is grateful for support from the UKRI Future Leaders Fellowship (grant MR/T042842/1). e-MERLIN is a National Facility operated by the University of Manchester at Jodrell Bank Observatory on behalf of STFC, part of UK Research and Innovation. MB acknowledges support from the ERC-Stg DRANOEL, no 714245. MJH acknowledges support from the UK Science and Techology Facilities Council (ST/R000905/1). JM acknowledges financial support from the State Agency for Research of the Spanish MCIU through the ``Center of Excellence Severo Ochoa'' award to the Instituto de Astrof\'isica de Andaluc\'ia (SEV-2017-0709) and from the grant RTI2018-096228-B-C31 (MICIU/FEDER, EU). This paper is based (in part) on results obtained with International LOFAR Telescope (ILT) equipment under project code LC14$\_$015. LOFAR (van Haarlem et al. 2013) is the Low Frequency Array designed and constructed by ASTRON. It has observing, data processing, and data storage facilities in several countries, that are owned by various parties (each with their own funding sources), and that are collectively operated by the ILT foundation under a joint scientific policy. The ILT resources have benefited from the following recent major funding sources: CNRS-INSU, Observatoire de Paris and Université d'Orléans, France; BMBF, MIWF-NRW, MPG, Germany; Science Foundation Ireland (SFI), Department of Business, Enterprise and Innovation (DBEI), Ireland; NWO, The Netherlands; The Science and Technology Facilities Council, UK7.
\end{acknowledgements}
  \bibliographystyle{aa} 
  \bibliography{References} 
\end{document}